\documentclass[journal]{IEEEtran}

\usepackage{xcolor,soul,framed}

\ifCLASSINFOpdf
   \usepackage[pdftex]{graphicx}
   \graphicspath{{../pdf/}{../jpeg/}}
   \DeclareGraphicsExtensions{.pdf,.jpeg,.png}
\else
   \usepackage[dvips]{graphicx}
   \graphicspath{{../eps/}}
   \DeclareGraphicsExtensions{.eps}
\fi
\usepackage{amsmath}
\usepackage{
algorithm,
algpseudocode,
amssymb
}

\usepackage{booktabs}

\ifCLASSOPTIONcompsoc
 \usepackage[caption=false,font=normalsize,labelfont=sf,textfont=sf]{subfig}
\else
 \usepackage[caption=false,font=footnotesize]{subfig}
\fi
\begin{document}
%
\title{Enriching Load Data Using Micro-PMUs and Smart Meters}
%
%
%


\author{Fankun Bu,~\IEEEmembership{Graduate Student Member,~IEEE,}
        Kaveh Dehghanpour,~\IEEEmembership{Member,~IEEE,}
        and Zhaoyu Wang,~\IEEEmembership{Member,~IEEE}
        
\thanks{This work was supported by Advanced Grid Modeling Program at the U.S. Department of Energy Office of Electricity under Grant DE-OE0000875. (\textit{Corresponding author: Zhaoyu Wang})}
\thanks{F. Bu, K. Dehghanpour and Z. Wang are with the Department of Electrical and Computer Engineering, Iowa State University, Ames, IA 50011
USA (e-mail: fbu@iastate.edu; kavehdeh1@gmail.com; wzy@iastate.edu).}
}

%
%

\markboth{}%
{Shell \MakeLowercase{\textit{et al.}}: Bare Demo of IEEEtran.cls for IEEE Journals}
%



\maketitle

\begin{abstract}
In modern distribution systems, load uncertainty can be fully captured by micro-PMUs, which can record high-resolution data; however, in practice, micro-PMUs are installed at limited locations in distribution networks due to budgetary constraints. In contrast, smart meters are widely deployed but can only measure relatively low-resolution energy consumption, which cannot sufficiently reflect the actual instantaneous load volatility within each sampling interval. In this paper, we have proposed a novel approach for enriching load data for service transformers that only have low-resolution smart meters. The key to our approach is to \textit{statistically} recover the high-resolution load data, which is masked by the low-resolution data, using trained probabilistic models of service transformers that have both high- and low-resolution data sources, i.e, micro-PMUs and smart meters. The overall framework consists of two steps: first, for the transformers with micro-PMUs, a Gaussian Process is leveraged to capture the relationship between the maximum/minimum load and average load within each low-resolution sampling interval of smart meters; a Markov chain model is employed to characterize the transition probability of known high-resolution load. Next, the trained models are used as \textit{teachers} for the transformers with only smart meters to decompose known low-resolution load data into targeted high-resolution load data. The enriched data can recover instantaneous load uncertainty and significantly enhance distribution system observability and situational awareness. We have verified the proposed approach using real high- and low-resolution load data. 
\end{abstract}


%
\IEEEpeerreviewmaketitle

\section{Introduction} \label{sec:intro}
%
%
%
%

\begin{figure}[t]
\centering
\includegraphics[width=0.85\linewidth]{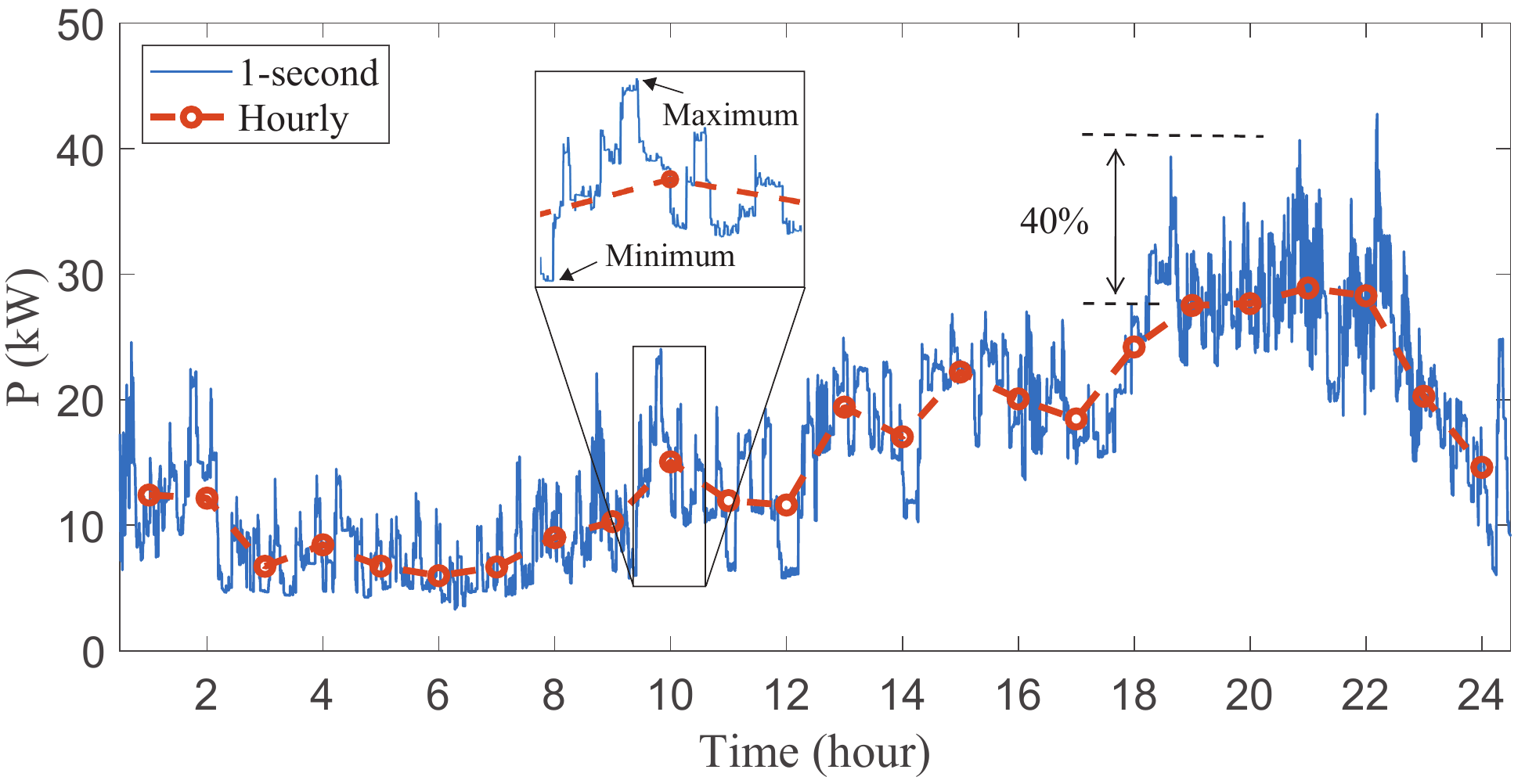}
\caption{A one-day real service transformer load curve with 1-second load data and corresponding hourly average load data.}
\label{fig:high_low_res_curves}
\end{figure}

\IEEEPARstart{A}{s} the advanced metering infrastructure (AMI) has been widely deployed in distribution systems in recent years, utilities have gained access to large amounts of smart meter (SM) data \cite{the_survey}. To take advantage of this data, which is both spatially and temporally fine-grained, researchers and industry practitioners have performed time-series power flow studies for optimizing network operation, expansion \cite{distribution_hand_book, hosting_capacity}, and integrating renewable energy resources \cite{ying_zhang}. In many cases, customer-level demands are aggregated to obtain service transformer-level loads for performing power flow studies \cite{distribution_hand_book,planning_reference_book}. However, the problem is that in most cases, SMs have a low sampling rate, e.g., one to four samples per hour. Thus, the average demand measured at such low resolutions cannot faithfully represent the uncertainties of the instantaneous load. As illustrated in Fig. \ref{fig:high_low_res_curves} for an exemplary transformer, the maximum 1-second load data has reached values 40\% times larger than the corresponding hourly SM reading within the same sampling interval. Also, compared to the hourly measurements, the instantaneous load shows a high level of variability, which has not been captured by the SMs. Therefore, recovering the \textit{masked} high-resolution load data is critical in enhancing distribution system situational awareness and granularity of modeling.

There is only a limited number of previous works focusing on load data enrichment. In \cite{xiangqi_zhu}, a top-down method is presented to generate service transformer-level high-resolution load profiles. First, low-resolution substation load profiles are allocated to service transformers via scaling. Then, the allocated profiles are decomposed into high-resolution load data by aggregating typical load patterns stored in variability and diversity libraries. In \cite{load_synthesis}, synthetic load datasets are created for four typical seasonal months using captured variability from high-resolution service transformer load data. To develop rich load data, researchers have added random noise to load data for modeling load uncertainty, as presented in \cite{add_noise}. In \cite{xiangqi_zhu_report}, a discrete wavelet transform (DWT)-based approach is proposed to parameterize intra-second variability of high-resolution transformer load data. To sum up, the primary limitations of previous load data enrichment methods are: the scaled substation load profiles allocated to service transformers differ from the actual load profiles since each transformer has a distinct load pattern \cite{Mingjian_Cui}, inaccuracy of adding random noise, and lack of specific methodology for applying the extracted load variability \cite{drawback_of_load_allocating}.

\begin{figure}
\centering
\includegraphics[width=0.9\linewidth]{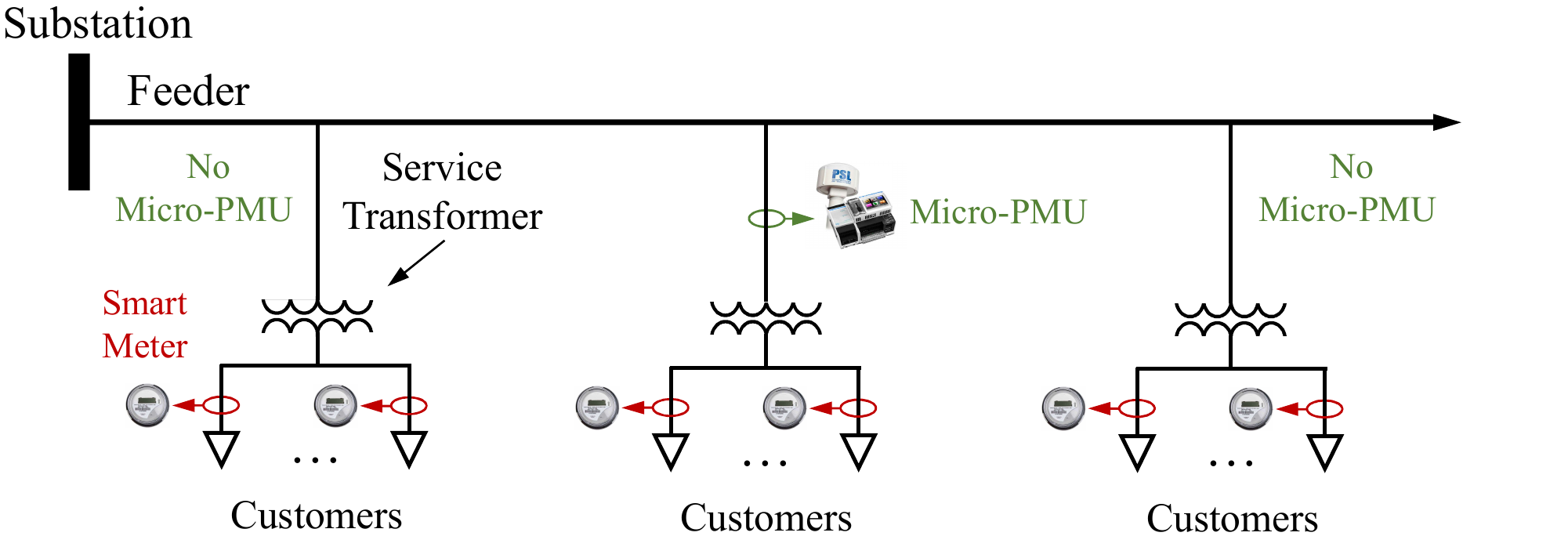}
\caption{Schematic diagram of a radial distribution feeder with diverse sensors.}
\label{fig:feeder_structure}
\end{figure}

\begin{figure}
\centering
\includegraphics[width=0.8\linewidth]{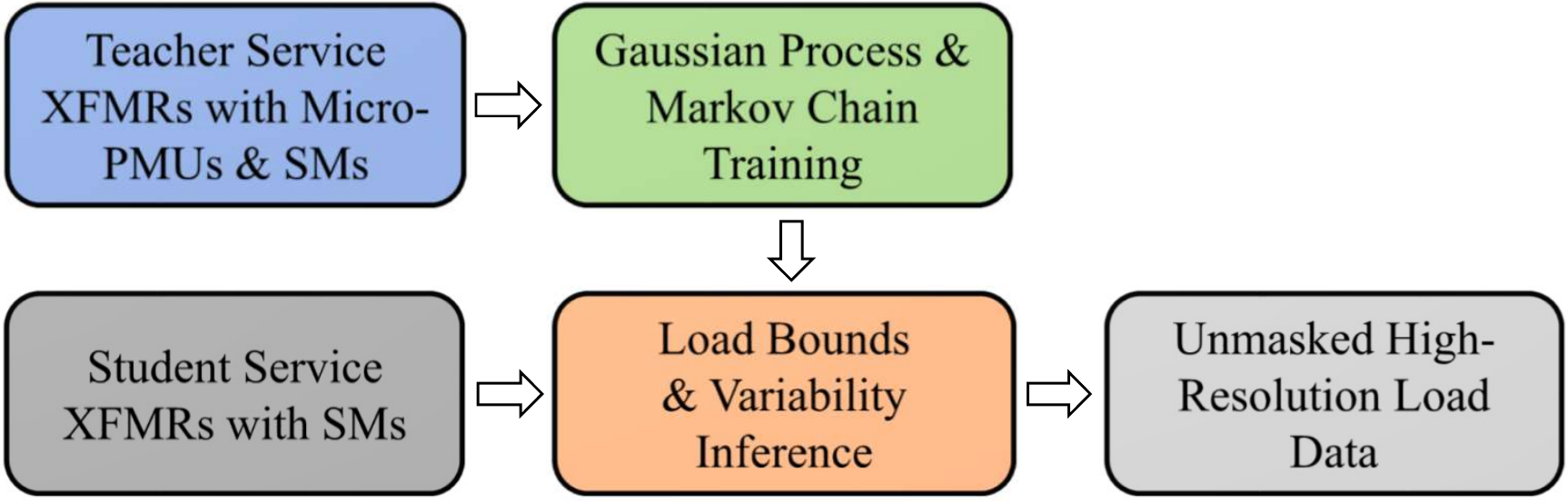}
\caption{Overall structure of the proposed load data enrichment approach.}
\label{fig:overall_structure}
\end{figure}

Considering the shortcomings of previous works, in this paper, we have developed a novel \textit{bottom-up} approach for enriching hourly load data for service transformers that only have SMs, by leveraging the high-resolution load data of service transformers with micro-PMUs and SMs. This concept is illustrated in Fig. \ref{fig:feeder_structure}, where the service transformer in the middle with rich load data is utilized to perform load data enrichment for the other two service transformers with only SMs. Before proceeding to specific steps, we have observed that each low-resolution load observation corresponds to a \textit{segment} of high-resolution load profile, as shown in Fig. \ref{fig:high_low_res_curves}. Therefore, enriching one known low-resolution load observation comes down to determining the maximum and minimum loads in the corresponding high-resolution load profile segment and inferring how the instantaneous load varies within those bounds. To do this, the proposed approach exploits learned probabilistic models that are trained using the high-resolution load data of service transformers \textit{with} micro-PMUs. Thus, the first stage is to train probabilistic models using known high-resolution load data of micro-PMUs. Specifically, a Gaussian Process is used to capture the relationship between the maximum/minimum bound and the average load. A Markov process is leveraged to model the probabilistic transition of instantaneous load within the bounds. These trained models for transformers with micro-PMUs form a \textit{teacher} repository. The second stage is to extend the trained probabilistic models to the service transformers that only have SMs, i.e., the \textit{students}, for enriching low-resolution load data. Specifically, the trained Gaussian Process models are employed to estimate the unknown maximum/minimum bound using the known low-resolution observation as the input, and the trained Markov models are used to probabilistically determine the variability of instantaneous load within the estimated maximum and minimum bounds. In addition, the load enrichment process in the second stage is performed using a weighted averaging operation, where the weights are determined by evaluating the similarity between low-resolution load data of the student and teacher transformers. The overall framework of our proposed approach is illustrated in Fig. \ref{fig:overall_structure}.

The primary contribution of our paper is that we have proposed a novel bottom-up inter-service-transformer load data enrichment approach using micro-PMUs and SMs. Our method takes full advantage of the fine-grained spatial and temporal granularity of SM and micro-PMU data. The rest of the paper is organized as follows: Section \ref{sec:GP_Markov_model} presents the process of training teacher models using data from transformers with micro-PMUs. Section \ref{sec:kmeans_inferring} describes the procedure of enriching load data for transformers with only SMs using the trained teacher models. In Section \ref{sec:casestudy}, case studies are analyzed, and Section \ref{sec:conclusion} concludes the paper.

\section{Constructing A Repository of Teacher Transformers}\label{sec:GP_Markov_model} 
The first step in load data enrichment is to train inference models based on high-resolution micro-PMU load data. In this section, inference model training includes two stages: load boundary inference model training, and load variability parameterization. Also, keep in mind that the inference model training process is performed for \textit{each} service transformer with a micro-PMU.
\subsection{Training Load Boundary Inference Model}\label{sec:GP} 
\begin{figure}[h]
\centering
\includegraphics[width=0.55\linewidth]{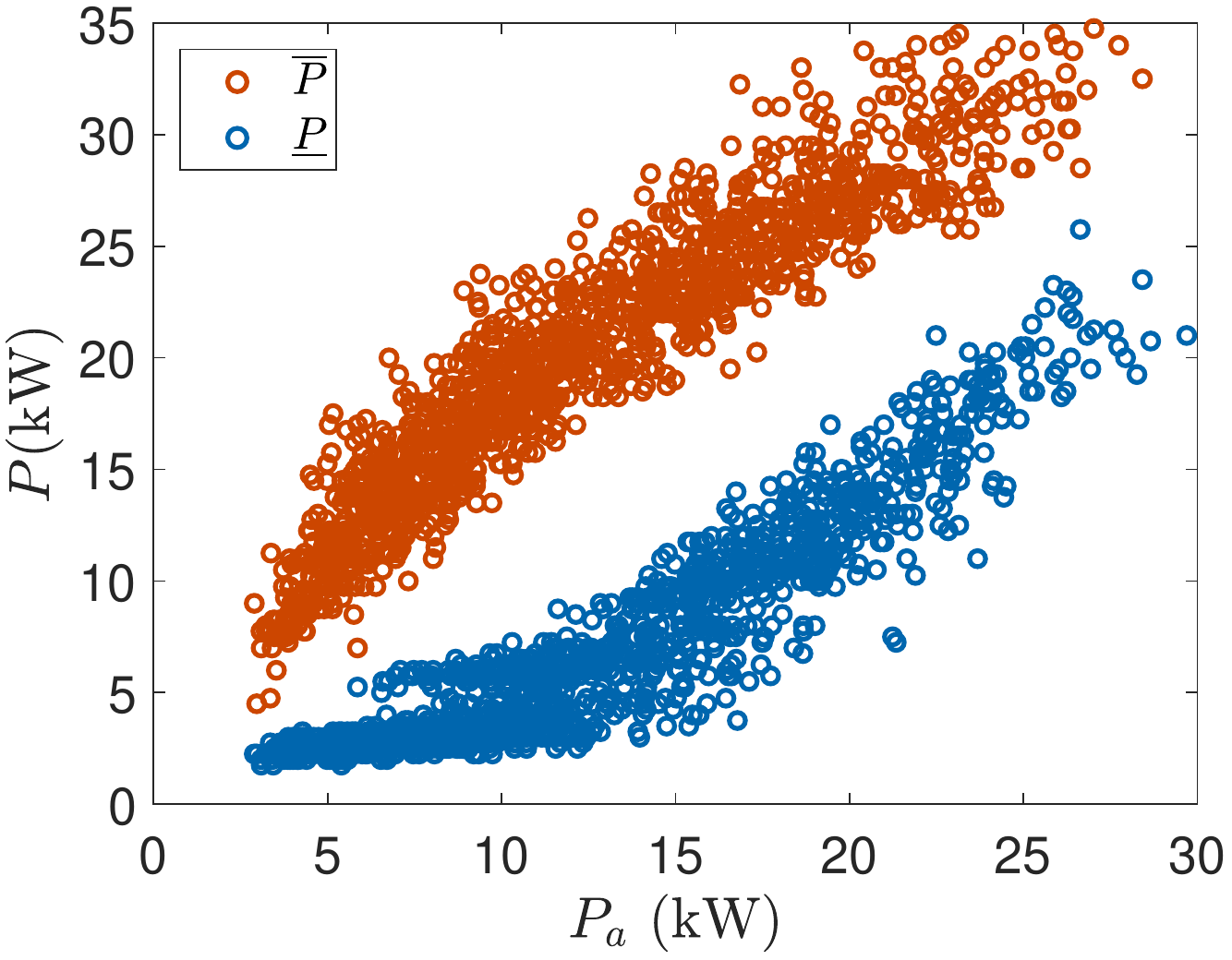}
\caption{Observation from real high-resolution load data for a service transformer.}
\label{fig:max_min_ave_P}
\end{figure}
Based on real high-resolution load data, we have observed that the average load over each low-resolution sampling interval, $P_a$, and the corresponding maximum/minimum load within that interval demonstrate a nonlinear relationship, as shown in Fig. \ref{fig:max_min_ave_P}. Note that $\overline{P}$ and $\underbar{$P$}$ denote the upper and lower bounds of instantaneous load within each sampling interval, respectively. Considering this, the Gaussian Process regression (GPR) technique, which shows satisfactory flexibility in capturing nonlinearity, is leveraged to train load boundary inference models \cite{pattern_recog_book}. 
The basic idea behind GPR is that if the distance between two explanatory variables is small, we have high confidence that the difference between corresponding dependent variables will be small as well. Specifically, using GPR, the upper bound of instantaneous load within the $t$'th hour, $\overline{P}(t)$, as a function of the hourly average load can be written as:
\begin{equation}  \label{eq:f_function}
\overline{P}(t) = f(P_a(t)),
\end{equation}
where, $P_a(t)$ denotes the average load over the $t$'th hour. Unlike deterministic approaches, where $f(P_a(t))$ is assumed to yield a single value for each $P_a(t)$, in GPR, $f(P_a(t))$ is a random variable. Intuitively, the distribution of $f(P_a(t))$ reflects the uncertainty of functions evaluated at $P_a(t)$. In GPR, the function $f(P_a(t))$ is distributed as a Gaussian process:
\begin{equation}  \label{eq:GPR_definition}
f \big(P_a(t) \big) \sim \mathcal{GP} \big(\mu(P_a(t)), K(P_a(t), P_a(t')) \big),
\end{equation}
where, $\mu(P_a(t))$ reflects the expected value of the maximum load inference function, and the covariance function $K(P_a(t), P_a(t'))$ represents the dependence between the maximum loads during different hour intervals. In our problem, the covariance function $K(\cdot,\cdot)$ is specified by the Squared Exponential Kernel function expressed as:
\begin{equation}  \label{eq:sqrt_exp_kernel}
K \big(P_a(t),P_a(t') \big) = \sigma_f^2 \text{exp} \bigg(-\frac{||P_a(t)-P_a(t')||_2^2}{2 \lambda^2}\bigg),
\end{equation}
where, $||\cdot||_2$ represents $l_2$-norm, $\sigma_f$ and $\lambda$ are hyper-parameters, which are determined using cross-validation. Intuitively, (\ref{eq:sqrt_exp_kernel}) measures the distance between $P_a(t)$ and $P_a(t')$, which can also reflect the similarity between $\overline{P}(t)$ and $\overline{P}(t')$, as shown in Fig. \ref{fig:max_min_ave_P}. 

For each service transformer with a micro-PMU, the average load and corresponding maximum load during each hour interval are known and provide a training dataset. Thus, applying (\ref{eq:GPR_definition}) to the entire training dataset consisting of $N$ hourly average and maximum load pairs, $\{(P_a(1),\overline{P}(1)),\cdots,(P_a(N),\overline{P}(N))\}$, an $N$-dimensional joint Gaussian distribution can be constructed as:
\begin{equation}   \label{eq:N_dim_Gau} 
\left[
\begin{array}{c}
f \big(P_a(1) \big)  \\
\vdots \\
f \big(P_a(N) \big)
\end{array}
\right] 
\sim \mathcal{N}
\Big(
\pmb{\mu}, \pmb{\Sigma}
\Big),
\end{equation}
where, 
\begin{subequations}  \label{eq:mu_sigma}
\begin{equation}
\pmb{\mu}=
\left[
\begin{array}{c}
\mu \big(P_a(1) \big)  \\
\vdots \\
\mu \big(P_a(N) \big)
\end{array}
\right],
\end{equation}    
\begin{equation}
\pmb{\Sigma}=
\left[
\begin{array}{ccc}
K\big(P_a(1),P_a(1) \big) & \cdots  & K\big(P_a(1),P_a(N) \big) \\
\vdots  & \ddots & \vdots\\
K\big(P_a(N),P_a(1) \big) & \cdots  &  K\big(P_a(N),P_a(N) \big)
\end{array}
\right].
\end{equation}
\end{subequations}
The joint Gaussian distribution formulated in (\ref{eq:N_dim_Gau}) represents a trained nonparametric \textit{maximum} load inference model. Also, the same procedure can be applied to the hourly average and minimum load pairs, $\{(P_a(1), \underbar{$P$}(1)), \cdots, (P_a(N), \underbar{$P$}(N)\}$, to obtain a trained nonparametric \textit{minimum} load inference model. 

In summary, for each service transformer with a micro-PMU, we can obtain two trained GPR models for inferring the maximum and minimum loads based on the corresponding hourly average load measured at the low-resolution sampling intervals.

\subsection{Training Load Variability Inference Model}\label{sec:Markov} 
\begin{figure}
\centering
\includegraphics[width=0.6\linewidth]{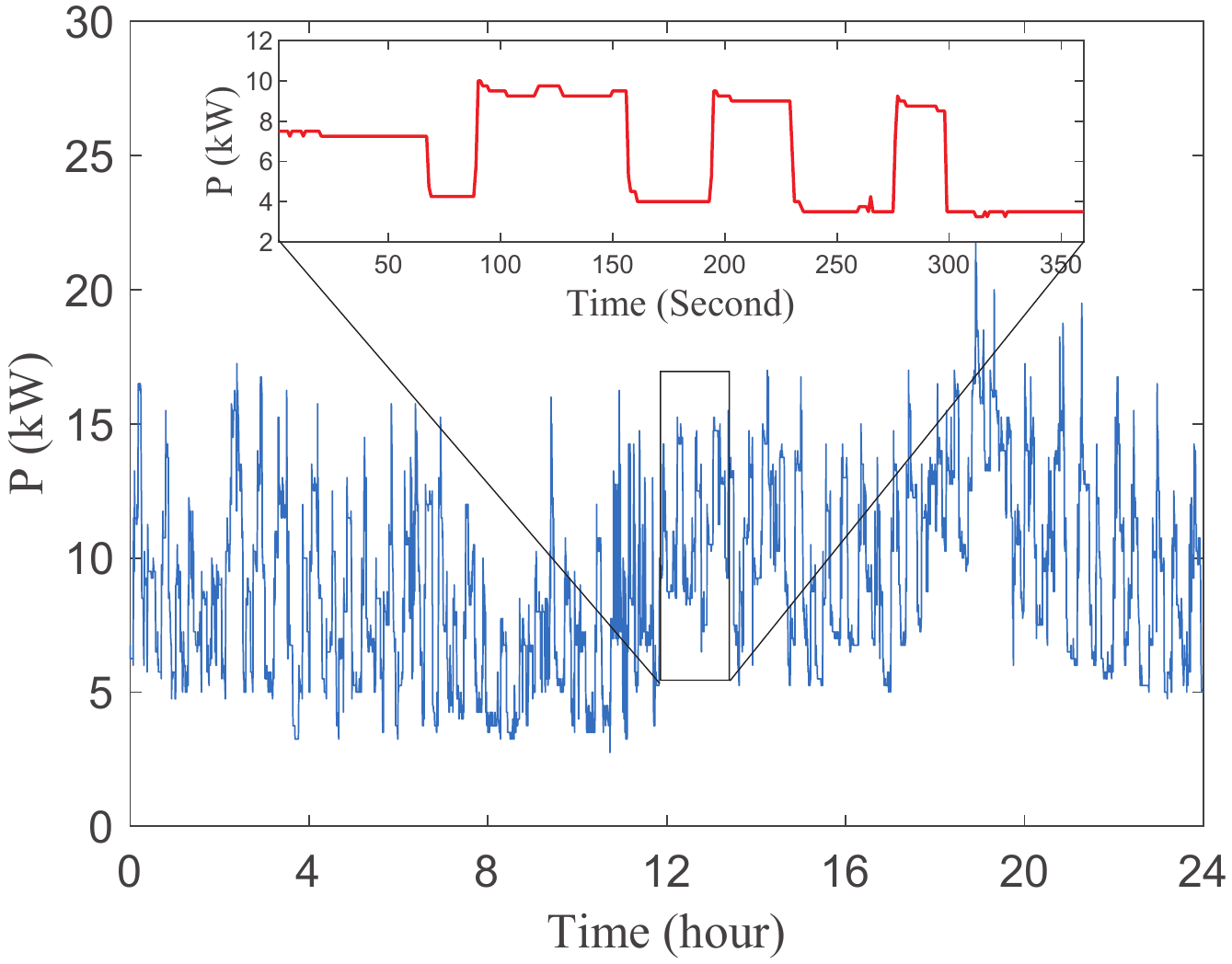}
\caption{Load variations within a day captured by high-resolution data.}
\label{fig:zoom_in_P_curve}
\end{figure}

Given an hourly average load observation, simply determining load boundaries is not sufficient for load data enrichment. We also have to learn how the load varies within these bounds. It is observed from real high-resolution load data that when an appliance is turned on, the load will jump to a certain level, as shown in Fig. \ref{fig:zoom_in_P_curve}. This process can be modeled as the Markov chain, which represents a system transitioning from one state to another over time. Also, it is observed from Fig. \ref{fig:zoom_in_P_curve} that once the load has transitioned to a certain level, it will stay almost invariant for a certain period of time. Therefore, the previous and current states can both affect the next state. Considering this, we have employed a second-order Markov model to capture the stochastic variability of the instantaneous load. Markov chains of second order are processes in which the next state depends on two preceding ones.

Since load is continuous, the first step to parameterize a Markov chain process is to discretize \textit{high-resolution} load measurements. Specifically, for the $i$'th high-resolution load observation during the $t$'th hour interval, the corresponding observed state is determined as:
\begin{multline}  \label{eq:discretization}
\quad \quad  S_t(i) = n_s, \quad n_s \in \{1,\cdots,N_s\}, t=1,\cdots,N,
\\ \text{if} \quad (n_s-1) \frac{\overline{P}(t)-\underbar{$P$}(t)}{N_s} \le P_t(i) - \underbar{$P$}(t) < n_s \frac{\overline{P}(t)-\underbar{$P$}(t)}{N_s},
\end{multline}
where, $N_s$ represents the total number of the unique discrete states and $P_t(i)$ is the $i$'th instantaneous load measurement during the $t$'th hour.

Also, it is observed from real high-resolution load data that different load levels display different stochastic processes. Typically, an air-conditioner cyclically starts and stops in the order of minutes. In contrast, the baseload, which is often caused by lighting and electronic devices, shows significantly longer cycles. In addition, the air-conditioning devices and baseload appliances show different average load levels over low-resolution sampling intervals due to different capacities. Therefore, to capture the different transition processes, the discretized observation states need to be divided into multiple subsets according to the hourly average load measurements. \textit{Each} subset is used to train a Markov chain model. Specifically, first, the entire collection of discretized observation states is split into $N_d$ subsets according to the corresponding low-resolution load observation, $P_a(t)$. The $j$'th subset is obtained as:
\begin{multline}  \label{eq:split_datasets}
\quad \pmb{D}_j=\{S_t(i)\}, \quad i \in \{1,\cdots,N'\}, t \in \{1,\cdots,N\},
\\ \text{if} \quad F\Big (\frac{(j-1)\times 100}{N_d} \Big) \le P_a(t) < F\Big (\frac{j\times 100}{N_d} \Big),  \quad
\end{multline}
where, $F(\cdot)$ is a function that returns percentiles of the entire set of low-resolution load observations, and $N'$ is the total number of discretized observation states in each low-resolution sampling interval. 

\begin{figure}
\centering
\includegraphics[width=0.85\linewidth]{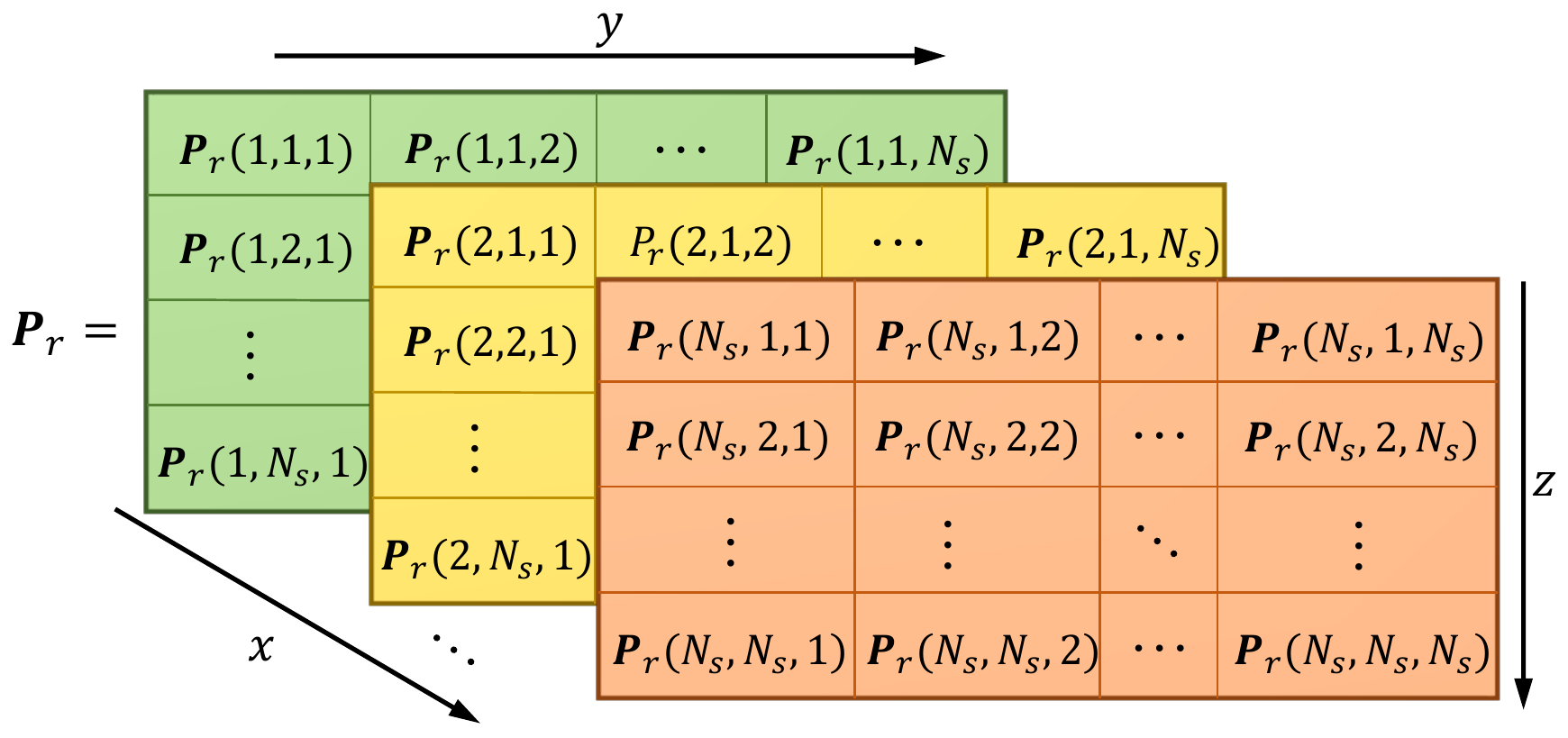}
\caption{Representation of the 3D probability transition matrix.}
\label{fig:3D_trans_matrix}
\end{figure}

Then, for each subset $\pmb{D}_j$, the stochastic process is parameterized by empirically estimating the transition probabilities between discrete observed states in terms of a transition matrix. A second-order Markov process consists of three states: the previous state, the current state, and the next state. Therefore, the stochastic transition matrix, $\pmb{P}_r$, is a three-dimensional (3D) array, as illustrated in Fig. \ref{fig:3D_trans_matrix}. Each element of $\pmb{P}_r$, $\pmb{P_r}(x,y,z)$, represents the probability of moving to state $z$ under the condition that the previous state is $x$ and the current state is $y$. For each subset $\pmb{D}_j$, elements of $\pmb{P}_r$ can be estimated from the frequencies of posterior states. Assume $\pmb{D}_j$ takes on the form of $\{S(i)\},i=1\cdots,N_s'$, where $N_s'$ is the total number of observation states in $\pmb{D}_j$, then the occurrence number at $(x,y,z)$ can be counted as:
\begin{multline}  \label{eq:counting}
\quad \quad \quad  \pmb{n}(x,y,z) = 
\displaystyle \sum_{i=2}^{N_s'-1}[S(i-1)==x \quad and  \\  S(i)==y \quad and \quad S(i+1)==z], \quad \quad 
\end{multline}
where, $[\cdot]$ is the Iverson bracket which converts any logical operation into 1, if the operation is satisfied, and 0 otherwise. ``$==$" stands for the ``equal to" operator and ``$and$" is the logical and operator. Thus, the elements of transition probability matrix are computed by:
\begin{equation}  \label{eq:transi_matr_element}
\pmb{P}_r(x,y,z)=\frac{\pmb{n}(x,y,z)}{\displaystyle \sum_{z}^{}\pmb{n}(x,y,z)}, \quad x,y=1,\cdots,N_s.
\end{equation}
For \textit{each} subset $\pmb{D}_j,j=1,\cdots,N_d$, (\ref{eq:transi_matr_element}) is performed to obtain a 3D stochastic transition matrix. Moreover, for \textit{each} service transformer with a micro-PMU, the entire above-mentioned procedure for parameterizing variability is conducted to obtain $N_d$ stochastic transition matrices.

\section{Enriching Load Data for Transformers with only Smart Meters}\label{sec:kmeans_inferring}  
\subsection{Determining Teacher Weights}\label{sec:Kmeans} 
Recall that our goal is to recover the high-resolution load data masked by the low-resolution load data. In this procedure, we leverage teacher models of transformers with micro-PMUs for service transformer with only SMs. Note that there might be more than one teacher transformer serving the same number of customers as the student transformer supplies. Different teacher transformers have different load behaviors. Thus, it is necessary to determine the learning weights corresponding to particular teacher transformers. These weights are determined by evaluating customer-level load similarity between the teacher and student transformers using low-resolution load data.

Specifically, for the $i$'th customer served by a student transformer, we can obtain a typical daily load pattern, $\pmb{P}_i$, which reflects customer behavior and the total capacity of appliances \cite{load_pattern}. Then, for a student transformer serving $N_c$ customers, we can obtain $N_c$ daily load patterns, $\{\pmb{P}_1,\cdots,\pmb{P}_{N_c}\}$. Similarly, for a teacher transformer supplying the same number of customers, we can also obtain $N_c$ daily load patterns. Since we have multiple teacher transformers, we can obtain a set of load pattern collections. The load pattern collection for the $k$'th teacher transformer is denoted by $\{\pmb{P}_1^k,\cdots,\pmb{P}_{N_c}^k\}, k=1,\cdots,N_t$, where, $N_t$ is the total number of teacher transformers. Then, load similarity between a student transformer and the $k$'th teacher transformer is evaluated as:
\begin{equation}  \label{eq:similarity}
W'_k = \frac{1}{N_c^2} \displaystyle \sum_{i=1}^{N_c}\displaystyle \sum_{j=1}^{N_c} ||\pmb{P}_i - \pmb{P}_j^k||, \quad k=1,\cdots,N_t.
\end{equation}
The $W'_k$'s in (\ref{eq:similarity}) are then normalized for a more convenient mathematical representation:
\begin{equation}  \label{eq:weight_normaliz}
W_k = \frac{W'_k}{\sum_{k=1}^{N_t}W'_k}.
\end{equation}

In summary, the normalized similarity weights reflect the confidence of a student transformer to learn from multiple teacher transformers for load data enrichment. 

\begin{figure}
\centering
\includegraphics[width=0.85\linewidth]{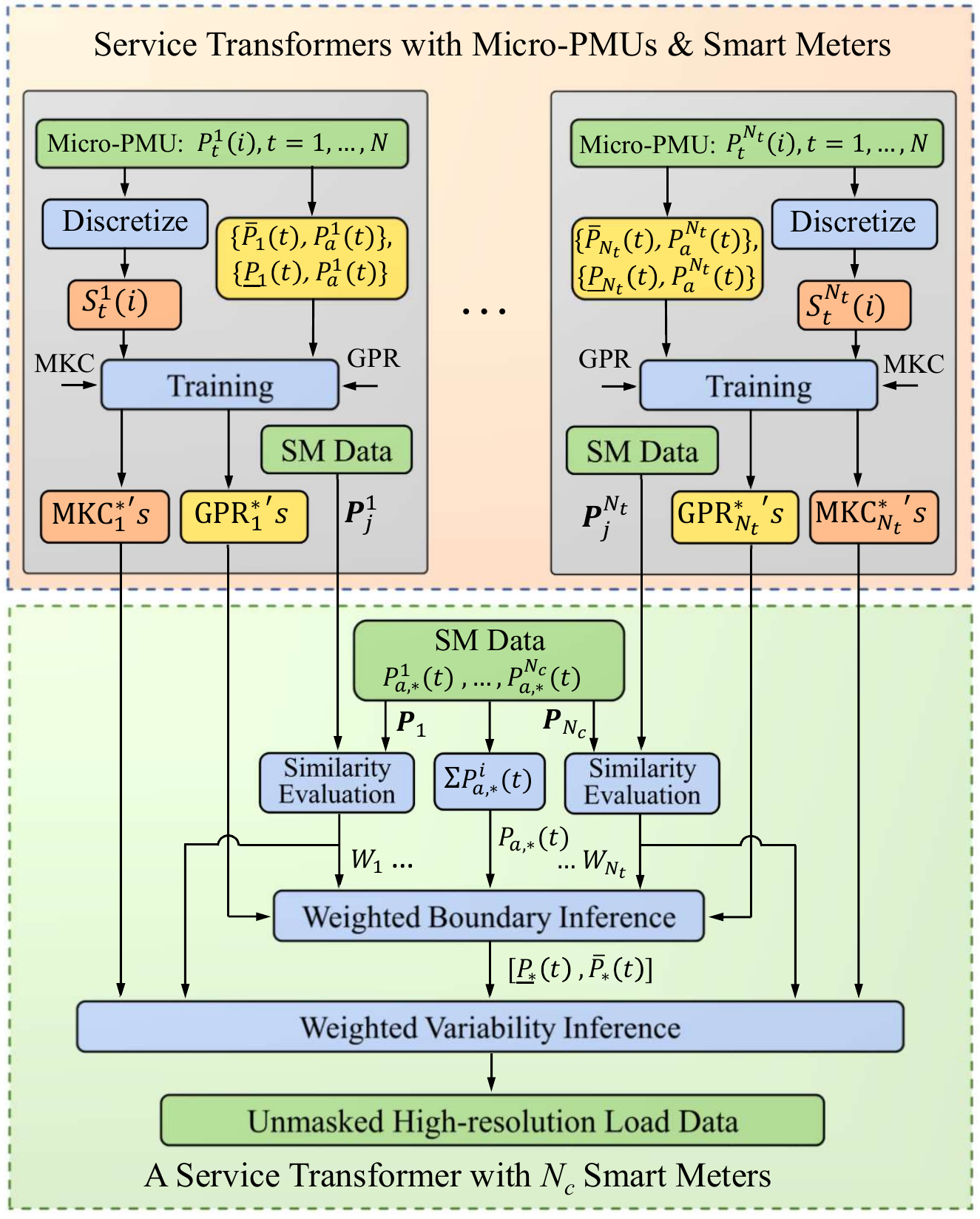}
\caption{Detailed steps of enriching load data.}
\label{fig:detailed_steps}
\end{figure}

\subsection{Enriching Load Data}\label{sec:Inferring} 
Using the normalized teacher weights, along with the load boundary and variability inference models derived in Section \ref{sec:GP_Markov_model}, we can conduct poor load data enrichment for service transformers that only have SMs. 
\subsubsection{Inferring Load Boundaries}\label{sec:Inferring_boundadry} 
In Section \ref{sec:GP}, for each teacher transformer with \textit{high-resolution} load data, we have obtained two GPR models for inferring the maximum and minimum loads given the corresponding average load over each low-resolution sampling interval. These two models are nonparametric and expressed in (\ref{eq:N_dim_Gau}). Specifically, the trained maximum load inference model for the $k$'th teacher transformer is expressed as:
\begin{equation}   \label{eq:N_k_dim_Gau} 
\left[
\begin{array}{c}
\overline{P}_k(1)  \\
\vdots \\
\overline{P}_k(N)
\end{array}
\right] 
=
\left[
\begin{array}{c}
f_k(P_{a,k}(1))  \\
\vdots \\
f_k(P_{a,k}(N))
\end{array}
\right] 
\sim \mathcal{N}
\Big(
\pmb{\mu}_k, \pmb{\Sigma}_k
\Big).
\end{equation}
To conduct load data enrichment, first, customer-level SM data are aggregated to obtain the load supplied by the student transformer, namely, $\{P_{a,*}(1), \cdots, P_{a,*}(N)\}$. Note that the transformer loss is approximated and added to the aggregate loads. Then, we assume the unknown upper bound of instantaneous load in terms of a function variable, $\overline{P}_{k,*}(t)=f_k(P_{a,*}(t)), t=1,\cdots,N$, follows a Gaussian distribution. By appending $\overline{P}_{k,*}(t)$ at the end of (\ref{eq:N_k_dim_Gau}), an ($N+1$)-dimensional Gaussian distribution can be formed as:
\begin{multline}\label{eq:N_k_dim_Gau_1} 
\quad \quad \quad \quad 
\left[
\begin{array}{c}
\overline{P}_k(1)  \\
\vdots \\
\overline{P}_k(N) \\
\overline{P}_{k,*}(t)
\end{array}
\right] 
=
\left[
\begin{array}{c}
f_k(P_{a,k}(1))  \\
\vdots \\
f_k(P_{a,k}(N)) \\
f_k(P_{a,*}(t))
\end{array}
\right] 
\\
\sim \mathcal{N}
\Big(
\left[
\begin{array}{c}
\pmb{\mu}_k\\
\pmb{\mu}_*
\end{array}
\right],
\left[
\begin{array}{cc}
\pmb{\Sigma}_k & \pmb{\Sigma}_{k*}\\
\pmb{\Sigma}_{k*}^T & \pmb{\Sigma}_{**}
\end{array}
\right]
\Big),
\quad \quad \quad 
\end{multline}   
where, $\pmb{\Sigma}_{k*}$ represents the training-test set covariances and $\pmb{\Sigma}_{**}$ is the test set covariance. In (\ref{eq:N_k_dim_Gau_1}), observations for $\{\overline{P}_k(1),\cdots,\overline{P}_k(N)\}$ are known and denoted by $\overline{\pmb{p}}_k=\{\overline{p}_k(1)),\cdots,\overline{p}_k(N))\}$. Thus, using the Bayes rule, the distribution of $\overline{P}_{k,*}(t)$ conditioned on $\overline{\pmb{p}}_k$ is obtained as:
\begin{equation}  \label{eq:inferred_load_distri}
\overline{P}_{k,*}(t) | \overline{\pmb{p}}_k \sim \mathcal{N} ({\mu}_*(t), \Sigma_*(t)),
\end{equation}
where, ${\mu}_*(t)=\pmb{\Sigma}_{k*}^T \pmb{\Sigma}_k^{-1} \overline{\pmb{p}}_k$ and $\Sigma_*(t)=\pmb{\Sigma}_{**}-\pmb{\Sigma}_{k*}^T \pmb{\Sigma}_k^{-1}\pmb{\Sigma}_{k*}$. Note that ${\mu}_*(t)$ denotes the most probable value of the estimated upper bound of instantaneous load given the average load during each low-resolution sampling interval.

Since we have $N_t$ teacher transformers, we can obtain a total of $N_t$ estimated maximum load candidates, namely, $\{\mu_*^1(t),\cdots,\mu_*^{N_t}(t)\}$. Also, considering load similarity between the student transformer and teacher transformers, a weighted-averaging operation is performed on all the inferred maximum loads to calculate a final estimated upper bound of instantaneous load:
\begin{equation}  \label{eq:inferred_final_load}
\overline{P}_*(t) = \displaystyle \sum_{k=1}^{N_t} W_k \mu_*^k(t), \quad t=1,\cdots,N.
\end{equation}

The same procedure introduced above is also applied to infer the unknown minimum load, $\underbar{$P$}_*(t)$, using the known average load over each low-resolution sampling interval. Once we have obtained the estimated load boundaries, then the trained probability matrices can be leveraged to infer load variability within each boundary.

\subsubsection{Inferring Load Variability}\label{sec:Inferring_variability} 
As introduced in Section \ref{sec:Markov}, each teacher transformer has $N_d$ extracted transition matrices corresponding to different load levels. Therefore, the first step in inferring the unknown high-resolution load variability is to determine which transition matrix to use. In other words, we need to find the variability inference matrix corresponding to the load level that the low-resolution load measurement belongs to. This is achieved by splitting the known low-resolution load observations of student transformer into $N_d$ subsets:
\begin{multline}  \label{eq:split_datasets_new}
\quad \pmb{P}_*^j=\{P_{a,*}(t)\}, \quad t \in \{1,\cdots,N\},j=1,\cdots,N_d,
\\ \text{if} \quad F\Big (\frac{(j-1)\times 100}{N_d} \Big) \le P_{a,*}(t) < F\Big (\frac{j\times 100}{N_d} \Big).
\end{multline}
Then, the $j$'th stochastic transition matrix of each teacher transformer is selected for enriching the low-resolution load measurements in the $j$'th subset of the student transformer, $\pmb{P}_*^j$. Moreover, considering that there is more than one teacher transformer, i.e., for each subset $\pmb{P}_*^j$, we have $N_t$ transition matrices to use. Thus, before proceeding to instantaneous load variability inference, a weighted averaging process similar to the load boundary estimation is conducted to obtain a comprehensive transition modal:
\begin{equation}  \label{eq:weightd_transi_model}
\pmb{P}_{r*}^j = \displaystyle \sum_{k=1}^{N_t} W_k \pmb{P}_{r}^{j,k}, \quad j=1,\cdots,N_d,
\end{equation}
where, $\pmb{P}_{r}^{j,k}$ stands for the transition matrix for the $k$'th teacher transformer based on the $j$'th subset of observation states, $\pmb{D}_j^k$. Then, for each low-resolution load measurement to be enriched, $P_{a,*}^j(t)$, the final targeted transition matrix, $\pmb{P}_{r*}^j$, and the inferred load boundary, $\{\overline{P}_*^j(t),\underbar{$P$}_*^j(t)\}$, are leveraged to generate the targeted high-resolution load data. Specifically, assume the previous state is $S_{t,*}^j(i-1)$, and the current state is $S_{t,*}^j(i)$, our goal is to determine the next state, $S_{t,*}^j(i+1)$, where, $i=1,\cdots,N'$, stands for the sequence number of states within the $t$'th low-resolution sampling interval. To do this, first, a random value at $i$, $U_*(i)$, is generated from the uniform distribution within the interval ($0,1$). Then, the state at ($i+1$) is determined by:
\begin{multline}  \label{eq:determine_state}
\quad \quad \quad \quad S_{t,*}^j(i+1) = z_*, \quad i=2,\cdots,N'-1,
\\ \text{if} \quad \displaystyle \sum_{z=1}^{z_*-1} \pmb{P}_{r*}^j \big(S_{t,*}^j(i-1),S_{t,*}^j(i),z \big) \le U_*(i) \\
< \sum_{z=1}^{z_*} \pmb{P}_{r*}^j \big(S_{t,*}^j(i-1),S_{t,*}^j(i),z \big). \quad \quad\quad
\end{multline}
Note that the generated $S_{t,*}^j(i)$'s are discrete state samples, therefore, they need to be transformed to specific load samples:
\begin{multline}  \label{eq:states_to_load}
P_{t,*}^j(i) =\underbar{$P$}_*^j(t) + \frac{S_{t,*}^j(i)\big(\overline{P}_*^j(t)-\underbar{$P$}_*^j(t) \big)}{N_s},
\quad i=1,\cdots,N'.
\end{multline}
Since there is more than one low-resolution time interval, the above procedure is conducted for \textit{each} low-resolution load observation. Also, since the low-resolution load observations are grouped into multiple subsets, the entire procedure introduced above is conducted through \textit{all subsets} of low-resolution load measurements. The detailed steps for load data enrichment for a student service transformer is illustrated in Fig. \ref{fig:detailed_steps}.

\section{Case Study}\label{sec:casestudy}
In this section, we have validated the proposed load data enrichment approach using real high- and low-resolution load data \cite{data_source}. 

\subsection{Dataset Description}
The dataset includes real 1-second load data for eight service transformers and hourly SM energy data for 185 customers. Among these customers, 36 are supplied by the 8 transformers with high-resolution load data (with micro-PMUs), and the remaining 149 customers are fed by the other 34 service transformers with low-resolution load data (with only SMs). To verify the performance of load data enrichment, the utility has also installed extra measuring devices to record 1-second load data for the service transformers with only SMs \cite{data_source}. The time range of the dataset is two months. In practice, micro-PMUs might have higher sampling rates than one sample per second, however, there is no fundamental difference in verifying the performance of our approach.

\subsection{Enriching Low-resolution Load Measurements}
\begin{figure}
\centering
\includegraphics[width=0.9\linewidth]{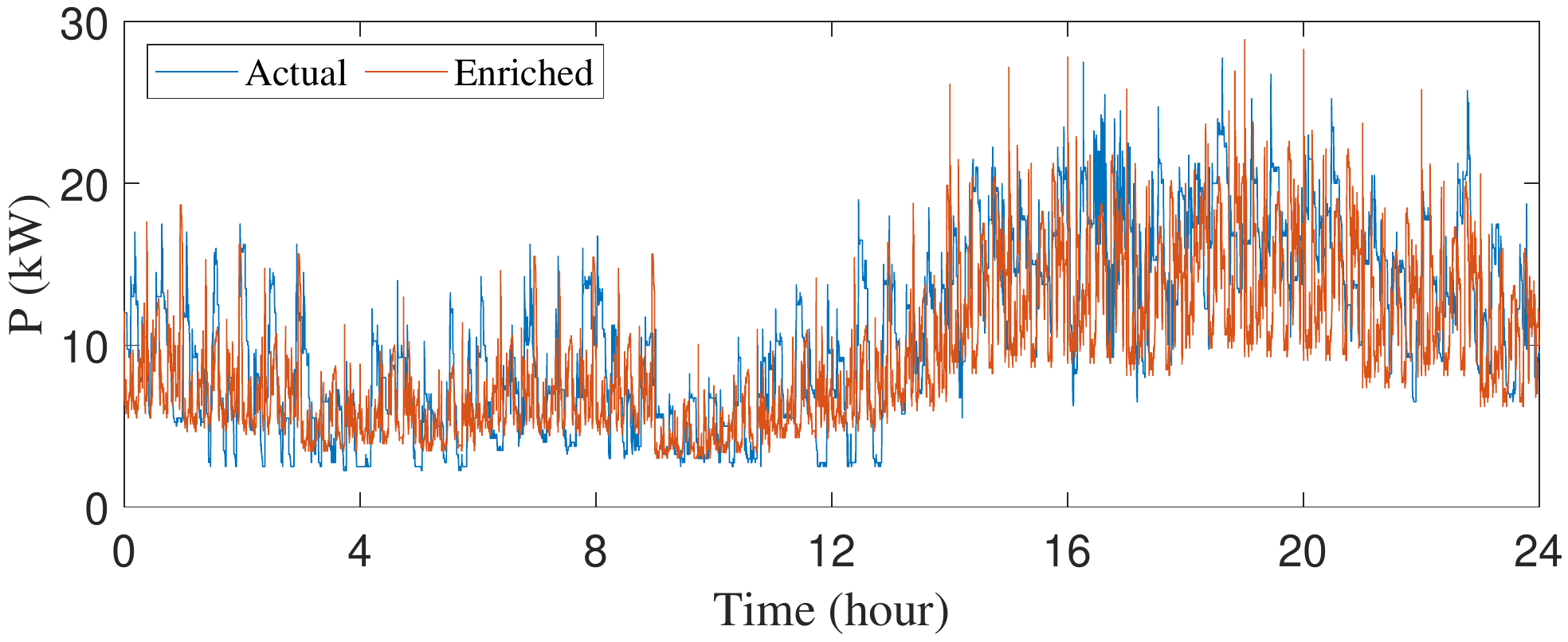}
\caption{One-day actual and enriched 1-second load data.}
\label{fig:actual_ebriched_Load}
\end{figure}

Fig. \ref{fig:actual_ebriched_Load} shows one-day actual and enriched 1-second load data for a service transformer. As can be seen, the enriched curve can accurately follow the actual basic load pattern. Note that our goal is not to force the enriched 1-second data to exactly track the actual load. Instead, our method is designed to restore the \textit{statistical} properties of instantaneous load given known low-resolution load observations obtained from hourly SM data.

\begin{figure}
\centering
\subfloat[Maximum\label{sfig:r_sqrt_max}]{
\includegraphics[width=0.42\linewidth]{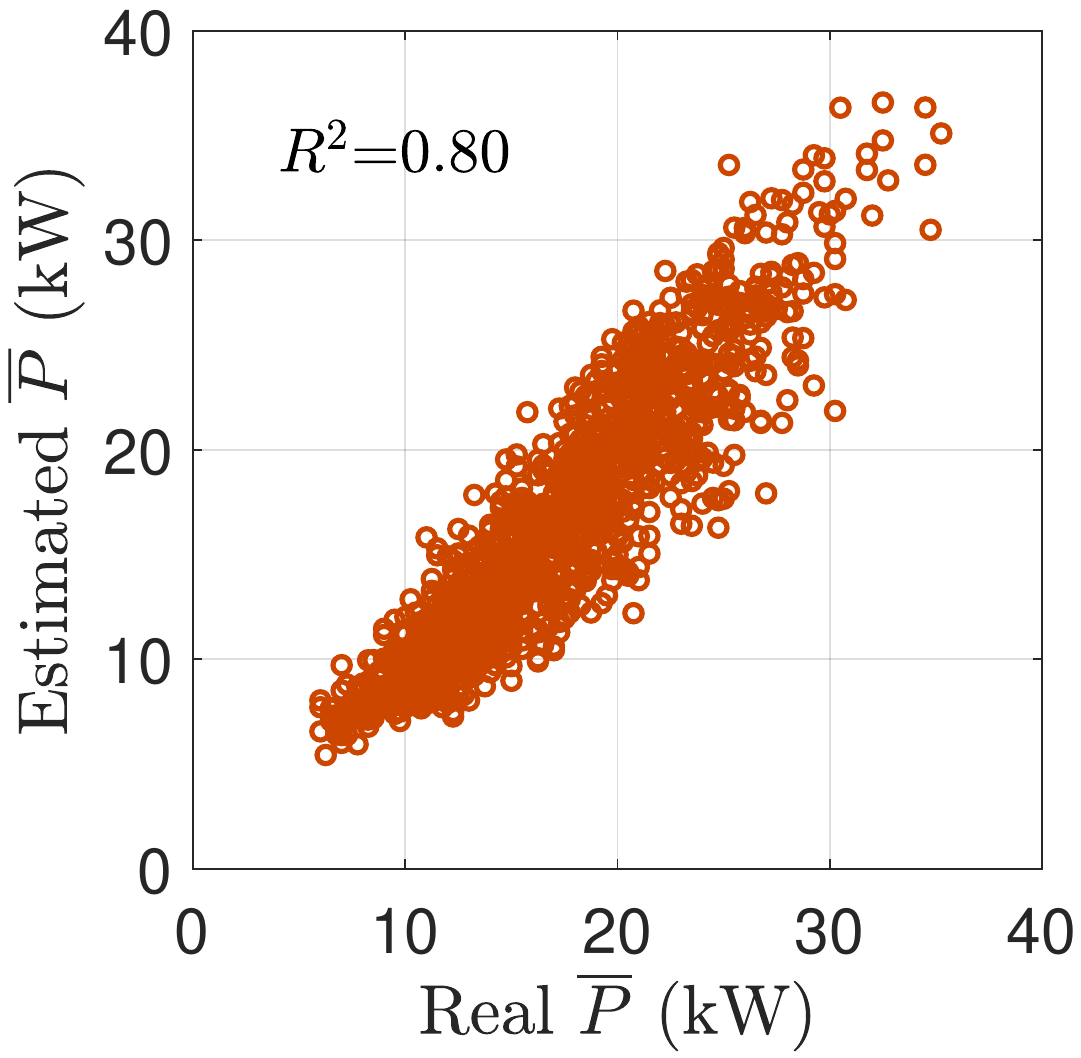}
}
\hfill
\subfloat[Minimum\label{sfig:r_sqrt_min}]{
\includegraphics[width=0.42\linewidth]{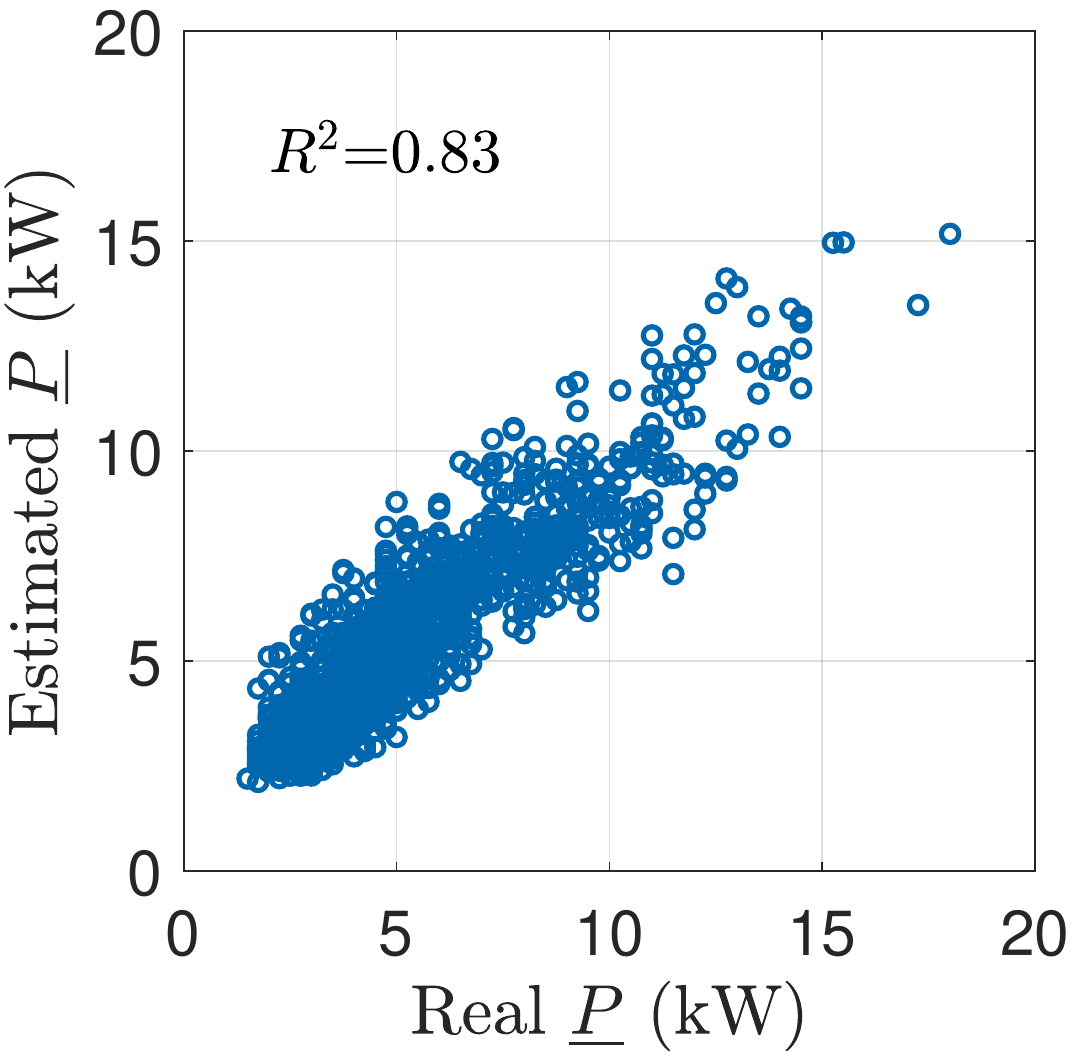}
}
\caption{The estimated maximum and minimum loads against the corresponding actual values.}
\label{fig:R_sqrt_values}
\end{figure}

One critical step of our proposed approach is to determine the masked maximum and minimum loads given a known average load observation on an hourly basis. Thus, it is of significance to examine the performance of the load boundary inference process. To do this, we have employed the coefficient of determination, $R^2$, for fitness evaluation, which is defined as:
\begin{equation}  \label{eq:R_squred}
R^2=1- \frac{\sum_{i=1}^{N}(y_i - \hat{y}_i)^2}{\sum_{i=1}^{N}(y_i - \overline{y})^2},
\end{equation}
where, $y_i$ denotes the real maximum/minimum instantaneous load, $\hat{y}_i$ denotes the corresponding inferred maximum/minimum instantaneous load, and $\overline{y}=\frac{1}{N}\sum_{i=1}^{N} y_i$. Fig. \ref{fig:R_sqrt_values} illustrates the effectiveness of load boundary estimation, and it can be seen that the estimated bound shows a linear relationship with the actual bound. The $R^2$ values for the upper and lower bounds are 0.80 and 0.83, respectively. This can also corroborate the accuracy of our proposed method.

\begin{figure}
\centering
\includegraphics[width=0.85\linewidth]{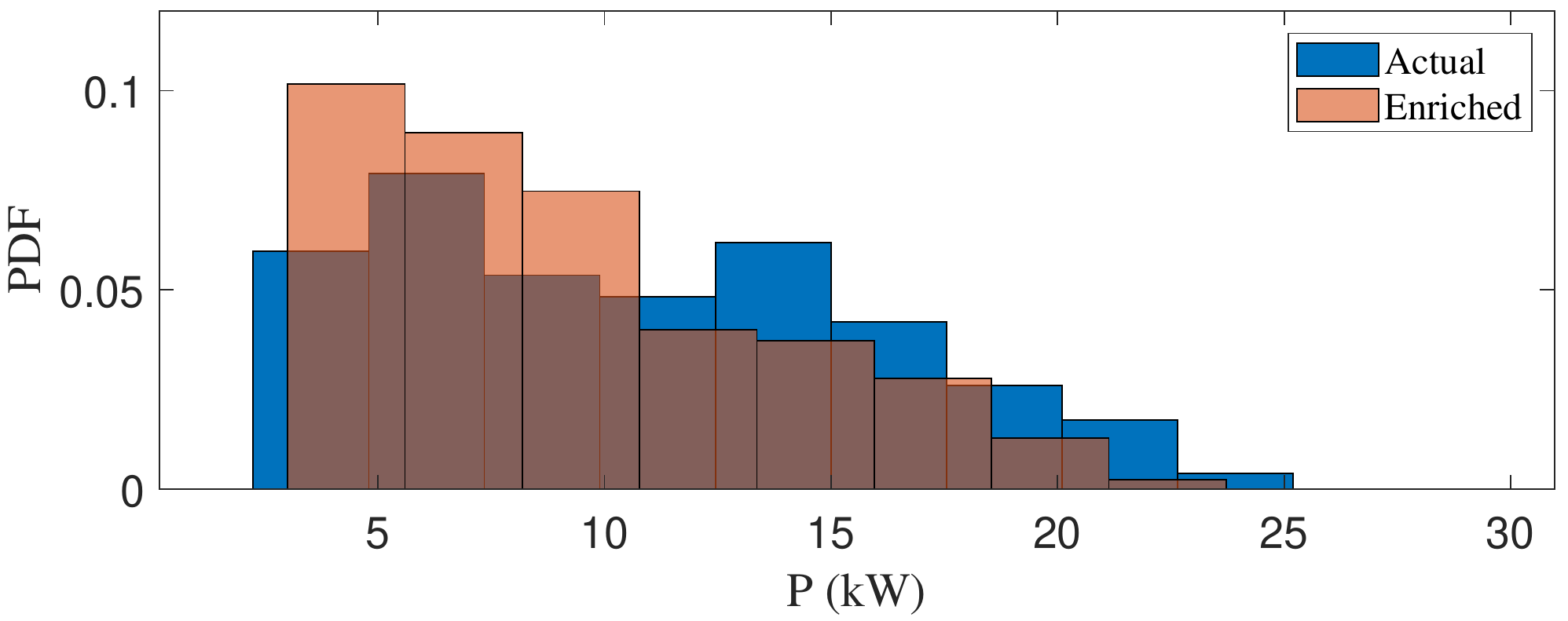}
\caption{Distributions of one-day actual and enriched 1-second load.}
\label{fig:histogram}
\end{figure}

\begin{figure}
\centering
\includegraphics[width=0.8\linewidth]{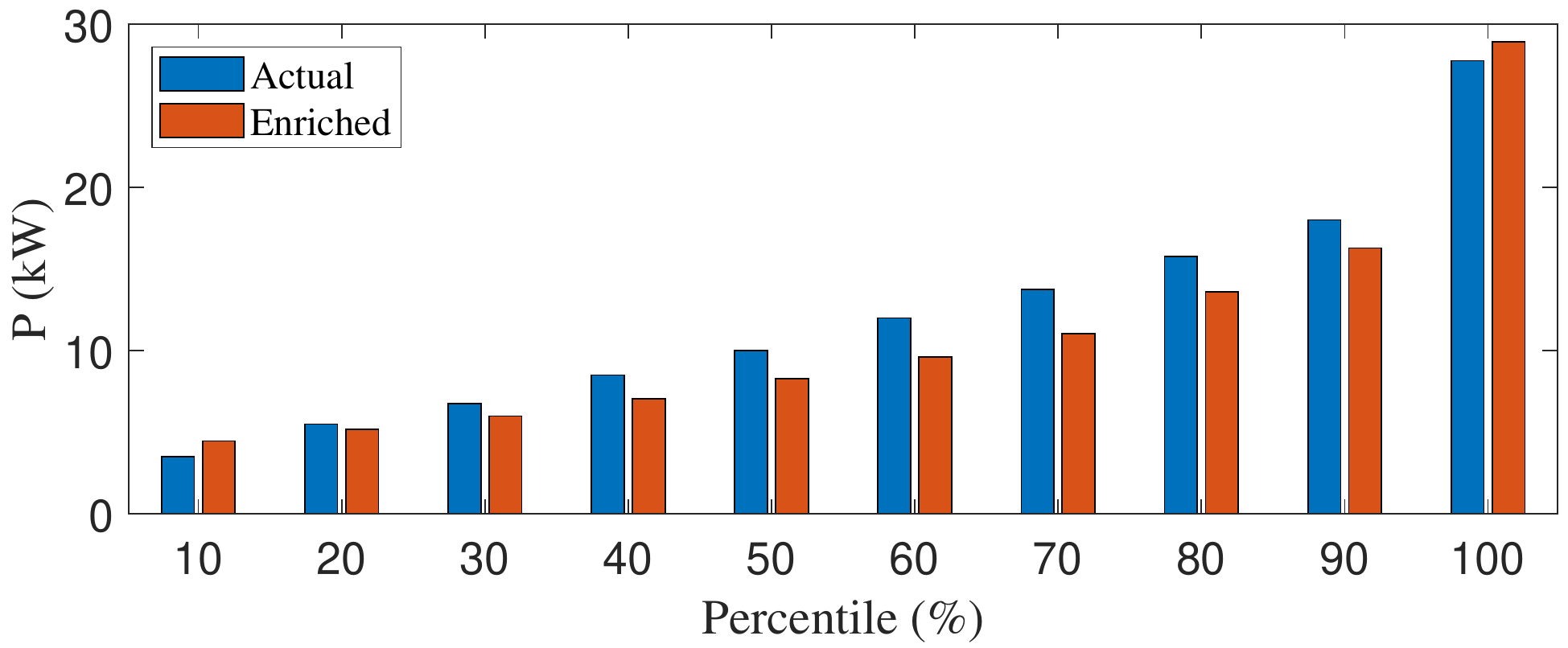}
\caption{One-day actual and enriched 1-second loads against percentiles.}
\label{fig:percentile_P}
\end{figure}

Note that our final goal is to recover the statistical properties of the high-resolution load within each low-resolution sampling interval. Therefore, the performance of our proposed approach needs to be evaluated  from the perspective of statistics. Fig. \ref{fig:histogram} illustrates the distributions of the actual and enriched load samples on the load curves shown in Fig. \ref{fig:actual_ebriched_Load}. It demonstrates that the enriched load distribution closely matches the actual load distribution. In comparison, the non-enriched load curve, which only includes 24 load observations, cannot sufficiently form a satisfactory distribution. In addition, to quantitatively assess load enrichment performance, we have examined the differences between the actual and enriched load values corresponding to different percentiles, as shown in Fig. \ref{fig:percentile_P}. It is demonstrated that the differences are significantly small, which also proves the effectiveness of our proposed approach from a statistical perspective. 

\subsection{Robustness to PV Integration}
\begin{figure}
\centering
\subfloat[Teacher transformers have PVs and student does not\label{sfig:teching_has_PV}]{
\includegraphics[width=0.75\linewidth]{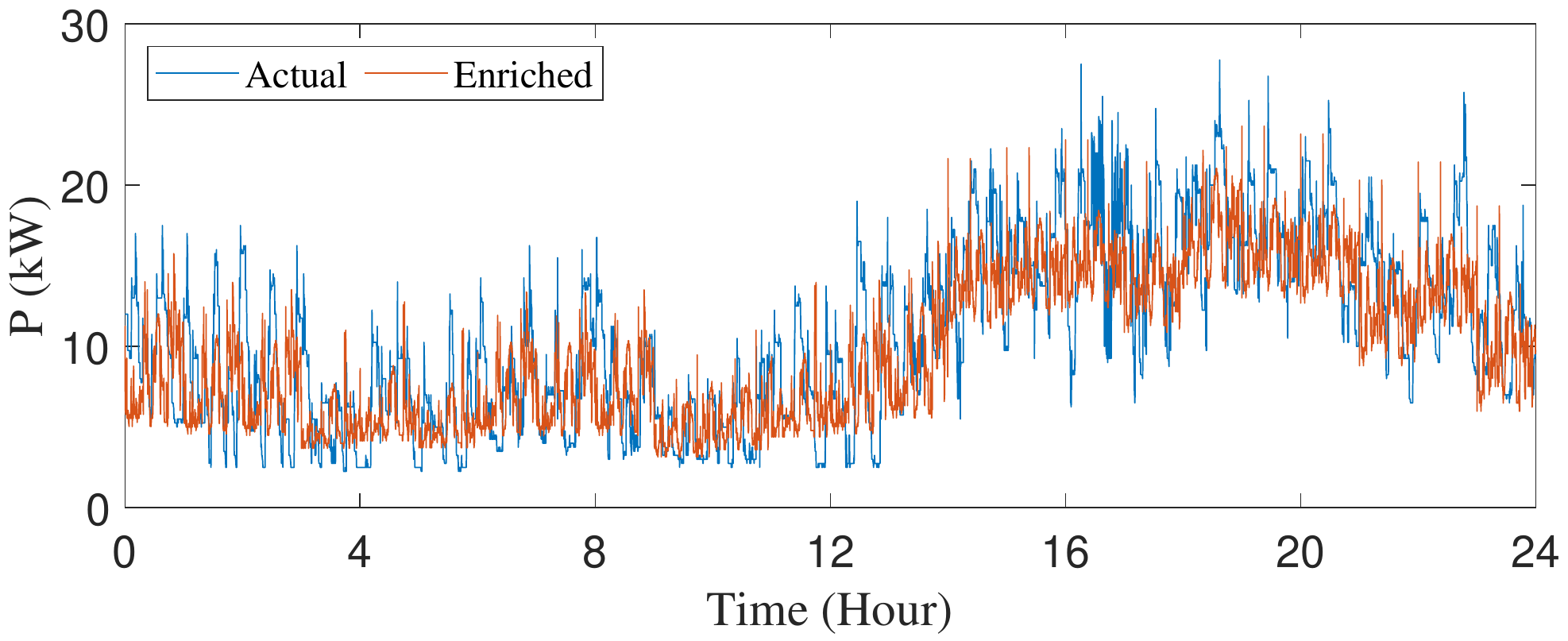}
}
\hfill
\subfloat[Teacher transformers do not have PVs and student does\label{sfig:student_has_PV}]{
\includegraphics[width=0.75\linewidth]{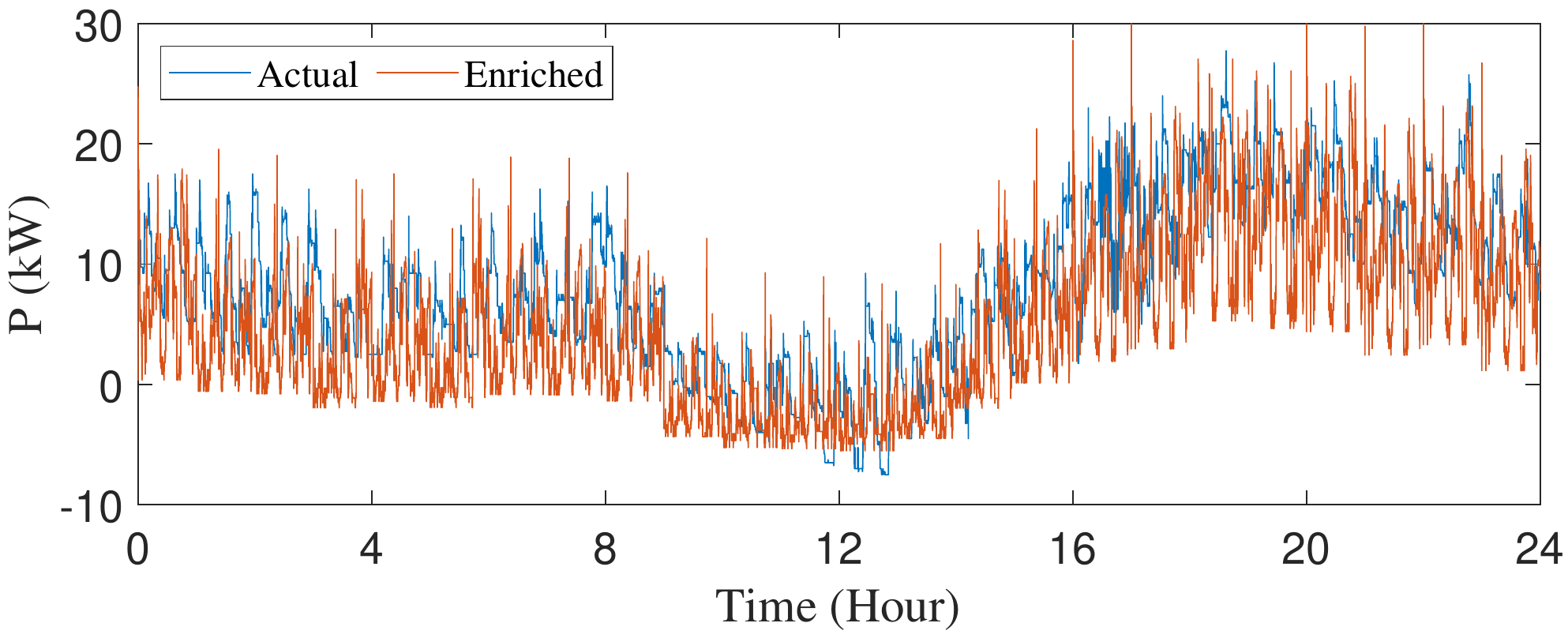}
}
\hfill
\subfloat[Both teacher and student transformers have PVs\label{sfig:Both_T_S_have_PV}]{
\includegraphics[width=0.75\linewidth]{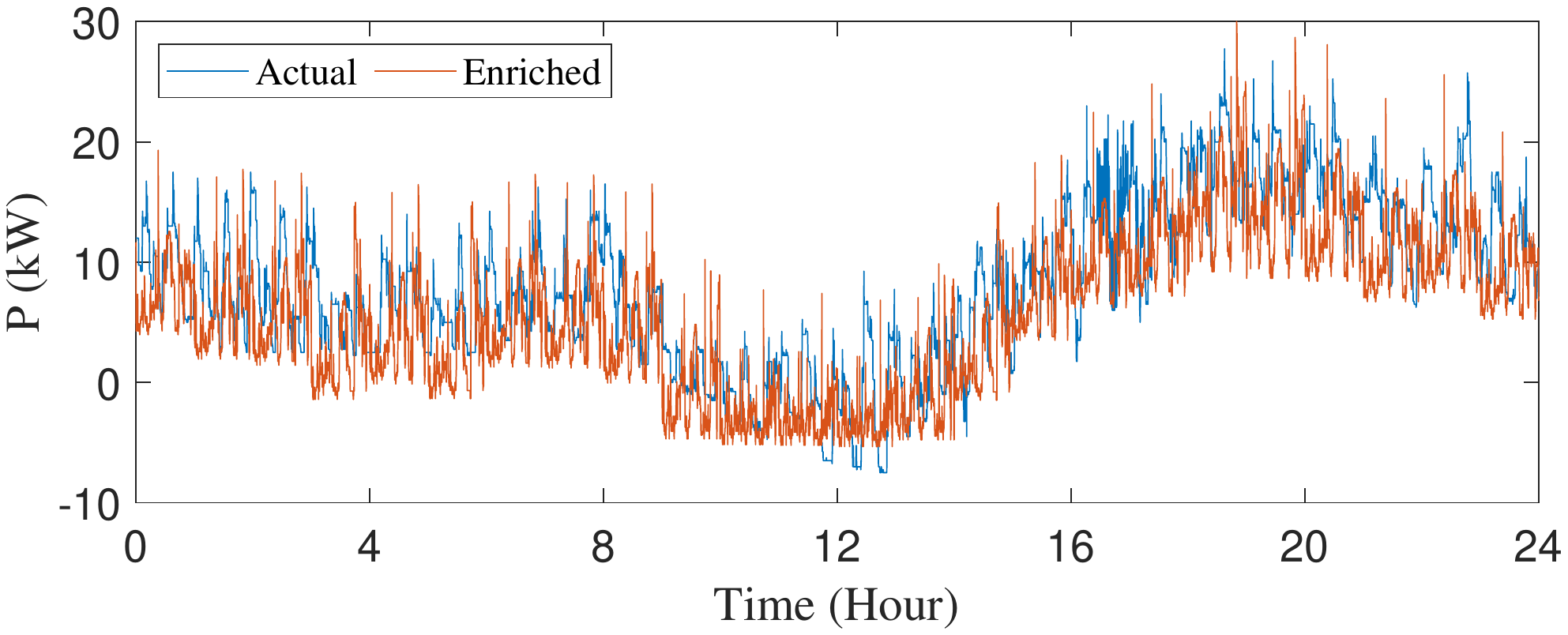}
}
\caption{Robustness of our proposed approach to small-scale PVs.}
\label{fig:PV_robustness}
\end{figure}

In modern distribution systems, PV integration is common for utilities. Therefore, it is necessary to test the performance of our load data enrichment approach under the condition of PV integration. Specifically, three scenarios are considered: the teacher transformers have installed customer-level PVs while the student transformer does not, the teacher transformers do not have PVs while the student transformer has, and both the teacher and student transformers have PVs. The enrichment results corresponding to the three cases mentioned above are shown in Fig. \ref{fig:PV_robustness}. It is demonstrated that the proposed approach can still achieve accurate high-resolution load data enrichment when the teacher and/or student transformers serve PV-installed customers.

\begin{figure}
\centering
\includegraphics[width=0.7\linewidth]{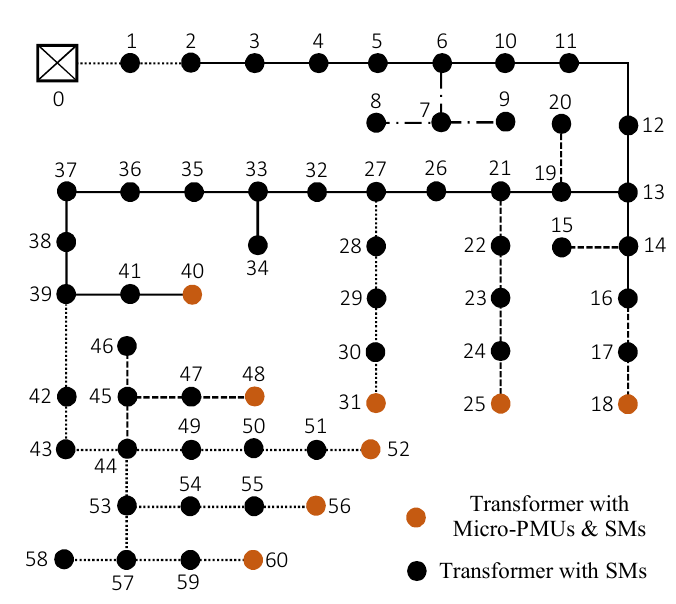}
\caption{One-line diagram of a real distribution system.}
\label{fig:Topology}
\end{figure}

\subsection{Performing Time-series Power Flow Studies}
To thoroughly examine the performance of our proposed approach, we have conducted time-series power flow studies by separately feeding the actual and enriched loads into a real distribution system \cite{Test_system}. The one-line topology of the real distribution system is shown in Fig. \ref{fig:Topology}. Bus voltages obtained from power flow analysis, which are critical to distribution system operators, are used to evaluate our proposed approach. Specifically, we compare the distributions of bus voltages and voltage ramps obtained from power flow studies based on the actual and enriched high-resolution load data, respectively. The reason for assessing voltage ramp is that voltage ramp is significant for renewable energy integration \cite{xiangqi_zhu}. The voltage ramp $\triangle V$ is defined as the difference between the current voltage value and the last voltage value.

\begin{figure}
\centering
\subfloat[Voltage\label{sfig:bus_votlage}]{
\includegraphics[width=0.48\linewidth]{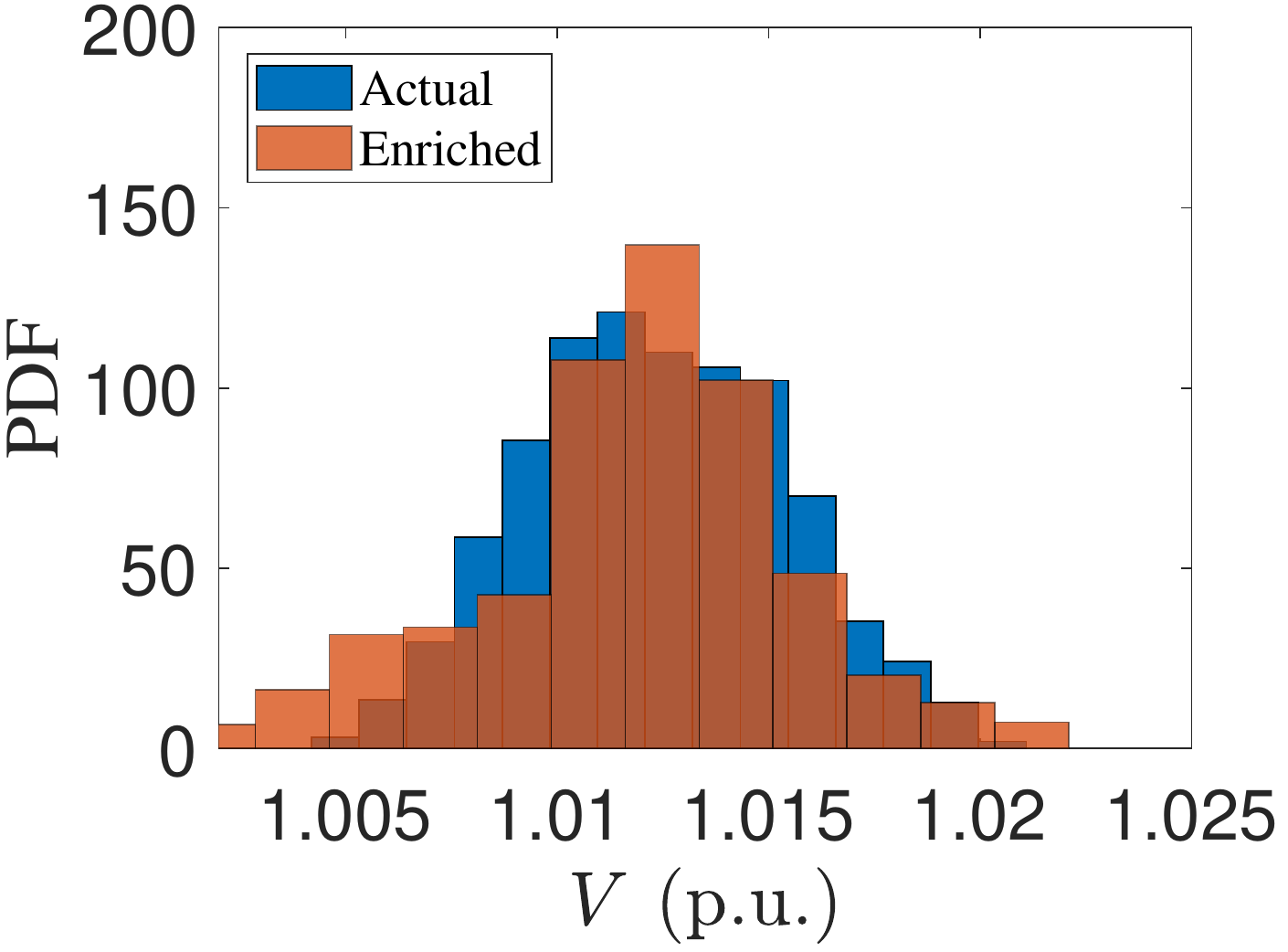}
}
\hfill
\subfloat[Voltage ramp\label{sfig:bus_voltage_ramp}]{
\includegraphics[width=0.46\linewidth]{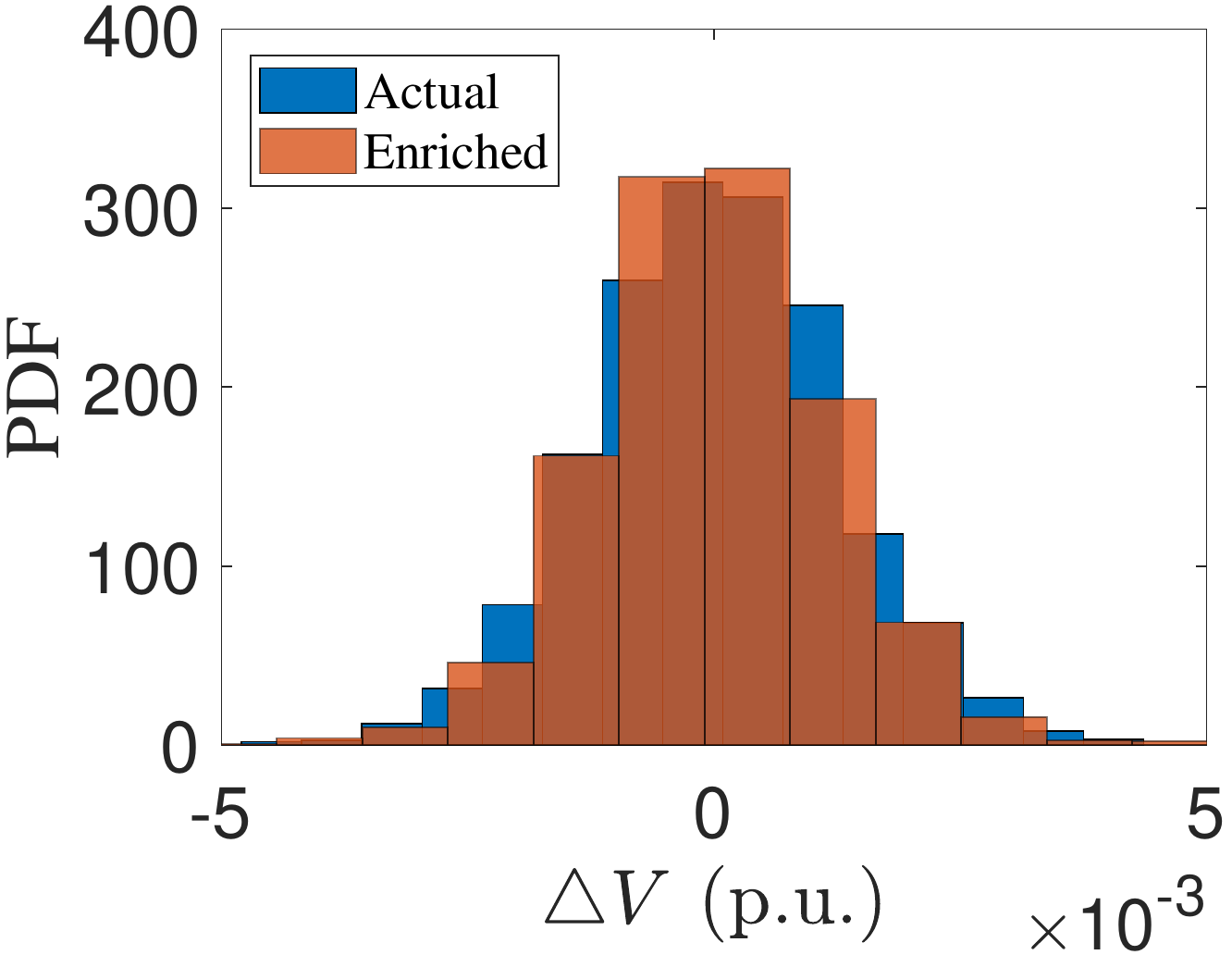}
}
\caption{Distributions of voltage and voltage ramp during a certain hour interval for Bus 57.}
\label{fig:voltage_distribution}
\end{figure}

Fig. \ref{fig:voltage_distribution} illustrates the distributions of voltages and voltage ramps during a certain hour interval for Bus 57 in the real distribution system. In Fig. \ref{sfig:bus_votlage}, it is observed that the empirical probability density function (PDF) of voltage based on the actual high-resolution load data can closely fit that based on the enriched high-resolution load data. For comparison, we have also performed a \textit{snapshot} flow study using the average load over the same hour interval. The per-unit voltage for Bus 57 is 1.0124, which is a single value without statistical properties. Therefore, the voltage distribution in Fig. \ref{sfig:bus_votlage} fully proves the capability of our proposed approach for recovering statistical characteristics masked by the low-resolution average load measurements. This capability can further enhance distribution system observability and situational awareness. A similar conclusion can be drawn for the voltage ramp, whose distribution is shown in Fig. \ref{sfig:bus_voltage_ramp}. As can be seen, the two voltage ramp distributions corresponding to the real rich load data and the enriched load data closely match each other. In comparison, the single bus voltage value based on the hourly average load cannot demonstrate probabilistic properties. It is important to point out that voltage distribution also depends on the specific structure of distribution systems in addition to specific load observations. For example, if a distribution system has very short line segments and a strong connection with a transmission system, then the bus voltage deviation might not be significant. In contrast, for a weak grid-connected distribution system with long line segments, the loads can have a strong impact on bus voltages.

\subsection{Performance Comparison}
\begin{figure}
\centering
\subfloat[Actual curve and enriched curve using our approach\label{sfig:our_approach}]{
\includegraphics[width=0.7\linewidth]{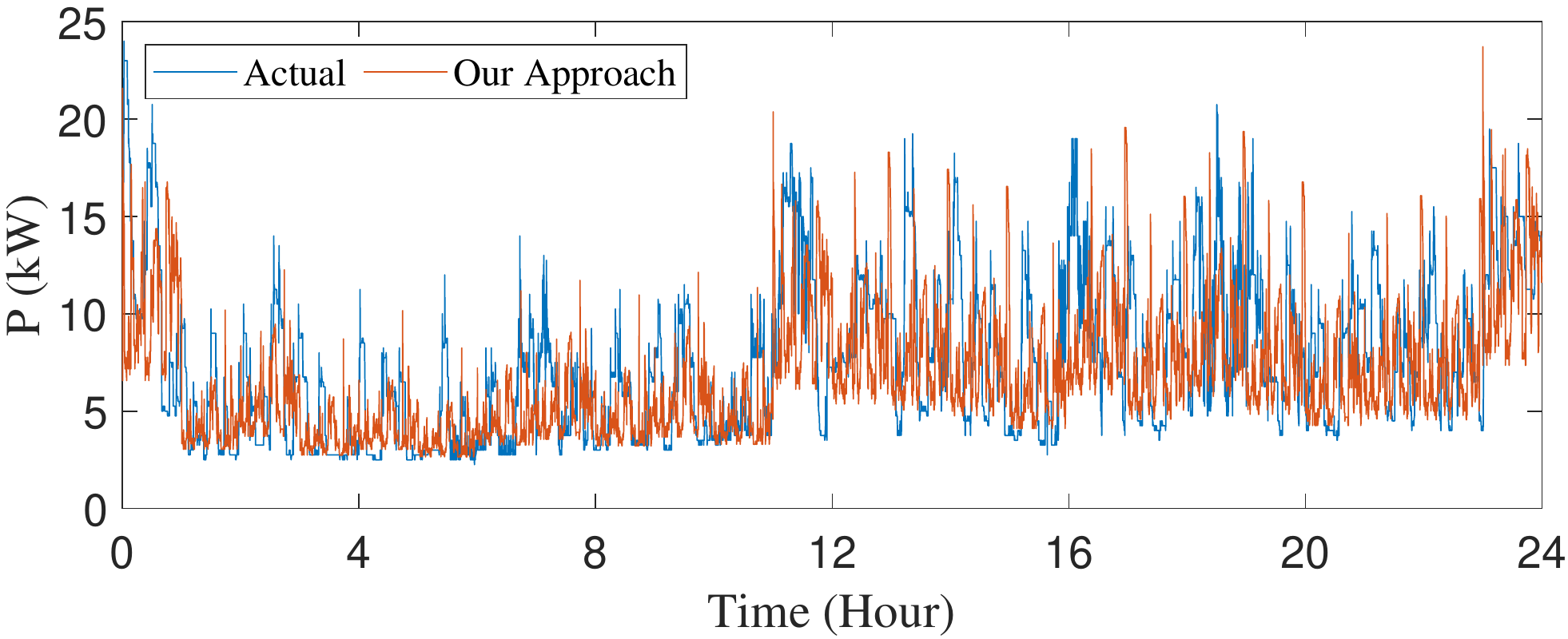}
}
\hfill
\subfloat[Actual curve and enriched curve using an allocation-based approach\label{sfig:xiangqi_zhu_approach}]{
\includegraphics[width=0.7\linewidth]{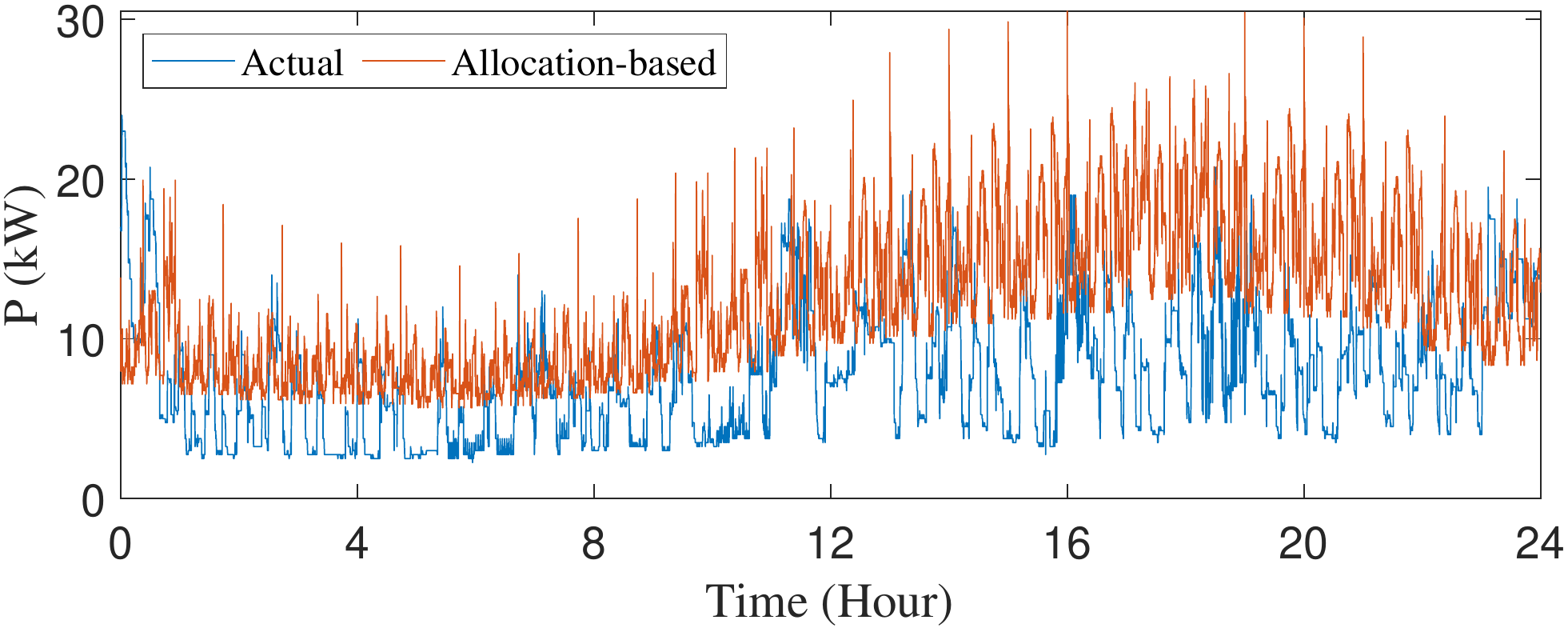}
}
\caption{Performance comparison between our approach and an allocation-based method.}
\label{fig:performance_comparison}
\end{figure}

It is of significance to compare the performance of our approach with other methods presented in previous works. We primarily focus on comparing our approach with an allocation-based methodology introduced in \cite{xiangqi_zhu}. The performance of these two approaches is shown in Fig. \ref{fig:performance_comparison}, where the actual and enriched load curves on a certain day are presented. In Fig. \ref{sfig:our_approach}, we can observe that the basic pattern of enriched 1-second load can flexibly follow the actual load variation, despite load uncertainty. This satisfactory performance results from two aspects, the fine spatial granularity of SM data and the design of load boundary inference process. In comparison, the allocation-based load enrichment approach cannot accurately track the basic load pattern, as demonstrated in Fig. \ref{sfig:xiangqi_zhu_approach}. The reason is that when allocating the substation load to service transformers, the allocated load follows a scaled substation-level load profile. However, this scaled load profile might not match the real load profile for a particular service transformer due to distinct load stochasticity.  

\section{Conclusion}\label{sec:conclusion}
This paper is devoted to temporally enriching low-resolution load data for service transformers that only have SMs, using high-resolution load data from service transformers with micro-PMUs and SMs. The entire process includes two stages, determining the maximum and minimum load bounds using known low-resolution load measurements and trained regression models, and inferring load variability within load boundaries using trained probabilistic transition models. The regression and transition models are trained using high-resolution load data from service transformers with micro-PMUs. We have used real high-resolution load data to prove that our approach is able to accurately recover high-resolution load data masked by the average load measurements over low-resolution sampling intervals. The enriched high-resolution load data can significantly enhance utilities’ grid-edge observability and situational awareness of distribution systems. Our paper's key findings are summarized as follows:
\begin{itemize}
\item The 1-second load within an hourly interval can be 40\% times larger or smaller than the corresponding average load during the same hour interval. By performing power flow studies, we have found that using the hourly average load for conducting power flow analysis cannot accurately capture the actual condition of distribution systems. Therefore, performing low-resolution power flow studies might cause significant errors, especially for those distribution networks that have a weak grid connection and long line segments.
\item The numerical experiments have verified that our proposed approach shows strong robustness and adaptability to PVs.
\item The numerical experiments have also demonstrated that our approach can accurately recover statistical properties of the instantaneous load within each low-resolution sampling interval of SM. The power flow studies show that our approach can faithfully reflect distribution system's  actual voltage conditions from a statistical perspective.
\end{itemize}

\ifCLASSOPTIONcaptionsoff
  \newpage
\fi



%



\bibliographystyle{IEEEtran}
\bibliography{IEEEabrv,ref}

\begin{thebibliography}{10}
\providecommand{\url}[1]{#1}
\csname url@rmstyle\endcsname
\providecommand{\newblock}{\relax}
\providecommand{\bibinfo}[2]{#2}
\providecommand\BIBentrySTDinterwordspacing{\spaceskip=0pt\relax}
\providecommand\BIBentryALTinterwordstretchfactor{4}
\providecommand\BIBentryALTinterwordspacing{\spaceskip=\fontdimen2\font plus
\BIBentryALTinterwordstretchfactor\fontdimen3\font minus
  \fontdimen4\font\relax}
\providecommand\BIBforeignlanguage[2]{{%
\expandafter\ifx\csname l@#1\endcsname\relax
\typeout{** WARNING: IEEEtran.bst: No hyphenation pattern has been}%
\typeout{** loaded for the language `#1'. Using the pattern for}%
\typeout{** the default language instead.}%
\else
\language=\csname l@#1\endcsname
\fi
#2}}

\bibitem{the_survey}
K.~Dehghanpour, Z.~Wang, J.~Wang, Y.~Yuan, and F.~Bu, ``A survey on state
  estimation techniques and challenges in smart distribution systems,''
  \emph{{IEEE} Trans. Smart Grid}, vol.~10, no.~2, pp. 2312--2322, Sep. 2018.

\bibitem{distribution_hand_book}
T.~A. Short, \emph{Electric Power Distribution Handbook}.\hskip 1em plus 0.5em
  minus 0.4em\relax Boca Raton, London, New York: CRC Press, 2014.

\bibitem{hosting_capacity}
F.~Ding and B.~Mather, ``On distributed {PV} hosting capacity estimation,
  sensitivity study and improvement,'' \emph{IEEE Trans. Sustain. Energy},
  vol.~8, no.~3, p. 1010–1020, Jul. 2017.

\bibitem{ying_zhang}
Y.~Zhang, J.~Wang, and Z.~Li, ``Uncertainty modeling of distributed energy
  resources: Techniques and challenges,'' \emph{Current Sustain. Energy Rep.},
  vol.~6, no.~2, pp. 42--51, Jun. 2019.

\bibitem{planning_reference_book}
H.~L. Willis, \emph{Power Distribution Planning Reference Book}.\hskip 1em plus
  0.5em minus 0.4em\relax New York: Marcel Dekker: CRC Press, 2004.

\bibitem{xiangqi_zhu}
X.~{Zhu} and B.~{Mather}, ``Data-driven distribution system load modeling for
  quasi-static time-series simulation,'' \emph{{IEEE} Trans. Smart Grid},
  vol.~11, no.~2, pp. 1556--1565, 2020.

\bibitem{load_synthesis}
M.~{Chamana} and B.~{Mather}, ``Variability extraction and synthesis via
  multi-resolution analysis using distribution transformer high-speed power
  data,'' in \emph{2017 19th International Conference on Intelligent System
  Application to Power Systems (ISAP)}, 2017, pp. 1--6.

\bibitem{add_noise}
A.~K. {Ghosh}, D.~L. {Lubkeman}, and R.~H. {Jones}, ``Load modeling for
  distribution circuit state estimation,'' \emph{IEEE Trans. Power Del.},
  vol.~12, no.~2, pp. 999--1005, 1997.

\bibitem{xiangqi_zhu_report}
X.~Zhu and B.~Mather, ``{DWT}-based aggregated load modeling and evaluation for
  quasi-static time-series simulation on distribution feeders preprint,'' Nat.
  Renew. Energy Lab., Golden, CO, USA, Tech. Rep. NREL/CP-5D00–70975, 2018.

\bibitem{Mingjian_Cui}
M.~{Cui}, J.~{Wang}, Y.~{Wang}, R.~{Diao}, and D.~{Shi}, ``Robust time-varying
  synthesis load modeling in distribution networks considering voltage
  disturbances,'' \emph{IEEE Trans. Power Syst.}, vol.~34, no.~6, pp.
  4438--4450, 2019.

\bibitem{drawback_of_load_allocating}
J.~{Peppanen}, C.~{Rocha}, J.~A. {Taylor}, and R.~C. {Dugan}, ``Secondary
  low-voltage circuit models—how good is good enough?'' \emph{IEEE Trans. Ind
  Appl.}, vol.~54, no.~1, pp. 150--159, 2018.

\bibitem{pattern_recog_book}
C.~M. Bishop, \emph{Pattern Recognition and Machine Learning}.\hskip 1em plus
  0.5em minus 0.4em\relax New York, NY, USA: Springer, 2009.

\bibitem{load_pattern}
Y.~{Yuan}, K.~{Dehghanpour}, F.~{Bu}, and Z.~{Wang}, ``A multi-timescale
  data-driven approach to enhance distribution system observability,''
  \emph{IEEE Trans. Power Syst.}, vol.~34, no.~4, pp. 3168--3177, 2019.

\bibitem{data_source}
K.~Nagasawa, C.~R. Upshaw, J.~D. Rhodes, C.~L. Holcomb, D.~A. Walling, and
  M.~E. Webber, ``{Data Management for a Large-Scale Smart Grid Demonstration
  Project in Austin, Texas},'' ser. Energy Sustainability, vol. ASME 2012 6th
  International Conference on Energy Sustainability, Parts A and B, Jul. 2012,
  pp. 1027--1031.

\bibitem{Test_system}
F.~Bu, Y.~Yuan, Z.~Wang, K.~Dehghanpour, and A.~Kimber, ``A time-series
  distribution test system based on real utility data,'' \emph{2019 North
  American Power Symposium}, pp. 1--6, Oct. 2019.

\end{thebibliography}

%








\end{document}